\documentclass[12pt]{article}
\usepackage{graphicx} 
\usepackage[a4paper, margin=1in]{geometry}
\usepackage[super,sort&compress]{natbib}
\usepackage{amsmath,amssymb,amsfonts}
\usepackage{booktabs}
\usepackage{longtable}
\usepackage{xcolor,subfig,lipsum}
\usepackage{newtxtext,newtxmath}
\usepackage{url}

\setcitestyle{super,comma} 

\newcommand{\btheta}{\mbox{\boldmath$\theta$}}
\newcommand{\bSigma}{\mbox{\boldmath$\Sigma$}}
\newcommand{\bpsi}{\mbox{\boldmath$\psi$}}
\newcommand{\bepsilon}{\mbox{\boldmath$\epsilon$}}
\newcommand{\balpha}{\mbox{\boldmath$\alpha$}}
\newcommand{\bbeta}{\mbox{\boldmath$\beta$}}
\newcommand{\bgamma}{\mbox{\boldmath$\gamma$}}

\newcommand{\bmu}{\mbox{\boldmath$\mu$}}
\newcommand{\bb}{\mbox{\boldmath$b$}}

\newcommand{\by}{\mbox{\boldmath$y$}}
\newcommand{\bZ}{\mbox{\boldmath$Z$}}
\newcommand{\bX}{\mbox{\boldmath$X$}}
\newcommand{\bW}{\mbox{\boldmath$W$}}
\newcommand{\bV}{\mbox{\boldmath$V$}}
\newcommand{\bY}{\mbox{\boldmath$Y$}}

\newcommand{\bI}{\mbox{\boldmath$I$}}

\title{Dynamic Prediction in Mixture Cure Models: \\ A Model-Based Landmarking Approach}
\author{Marta Cipriani$^1$ \and Marco Alfò$^1$ \and Mirko Signorelli$^2$}
\date{%
\small
    $^1$\emph{Department of Statistical Sciences, Sapienza University of Rome (IT)}\\%
    $^2$\emph{Mathematical Institute, Leiden University (NL)}%
}

\begin{document}

\maketitle

\begin{abstract}
Mixture cure models are widely used in survival analysis when a portion of patients is considered cured and is no longer at risk for the event of interest. In clinical settings, dynamic survival prediction is particularly important to refine prognosis by incorporating updated patient information over time. Landmarking methods have emerged as a flexible approach for this purpose, as they allow to summarize longitudinal covariates up to a given landmark time and to use these summaries in subsequent prediction. For mixture cure models, the only landmarking strategy available in the literature relies on the last observation carried forward (LOCF) method to summarize longitudinal dynamics up to the landmark time. However, LOCF discards most of the longitudinal information, does not correct for measurement error, and may rely on outdated values if observation times are far apart.
To overcome these limitations, we propose a sequential approach that integrates model-based landmarking within a mixture cure model. Initially, longitudinal covariates are modeled using (generalized) linear mixed models, from which individual-specific random effects are predicted. The predicted random effects are then incorporated as covariates into a Cox proportional hazards cure model. \\
We investigated the performance of the proposed approach under different cure fractions, sample sizes, and longitudinal data structures through an extensive simulation study. The results show that the model-based strategy provides more refined predictions compared to LOCF, even when the model is misspecified in favour of the LOCF approach. Finally, we illustrate our method using a real-world dataset on renal transplant patients.
\end{abstract}


\section{Introduction}
Dynamic prediction models are crucial tools in clinical research, as they leverage baseline and follow-up data to provide updated survival predictions as new information becomes available. This ability to adapt predictions over time is particularly valuable in personalised medicine, where therapeutic strategies can be tailored to the evolving profile of each patient.

Several approaches have been developed to implement dynamic prediction. The time-dependent Cox model\citep{therneau}, for example, incorporates time-varying covariates directly into the hazard function, offering a natural extension of the classical proportional hazards framework. Joint models provide a more integrated solution by simultaneously modelling the longitudinal trajectory of covariates and the survival process, typically by including individual-specific random effects that link the two submodels \citep{joint1, joint2}. An alternative and flexible approach is landmarking, which builds upon prediction models at pre-specified time points, known as landmark times, using only individuals still at risk at those times \citep{landmark}. In landmarking, repeated measures of time-dependent covariates observed up to the landmark time are first summarised, either through simple synthetic functions or statistical models, and then used as predictors in survival models alongside baseline covariates.

The way in which longitudinal information is summarised clearly plays a central role in determining the performance of these approaches. Landmarking methods can be broadly categorized into \textit{traditional} approaches and \textit{model-based} approaches. Traditional methods\cite{landmark} are based on simple strategies, such as those based on the last observation carried forward (LOCF) or the average of previous measurements. In contrast, model-based approaches employ more sophisticated statistical tools to capture the underlying structure of longitudinal data. These include linear mixed-effects models (LMM) and multivariate functional principal component analysis (MFPCA), both of which allow rich summaries of the time-varying covariates. Such model-based summaries have been used in combination with several prediction models, including MFPCA with Cox models \citep{li, gomon}, MFPCA with random survival forests \citep{lin}, LMMs with random survival forests \citep{devaux}, and LMMs with penalized Cox models \citep{signorelli, pencal}.

As interest in prediction methods increases, it has also become evident that the assumption of homogeneity in survival models is often violated. In such cases, cure models offer a valuable framework by explicitly accounting for a subgroup of people who are effectively \textit{cured} and not at risk of experiencing the event of interest, even with long-term follow-up\cite{amico}. The two most established approaches in this framework are the mixture cure model (MCM) \citep{boag, berkson} and the promotion time cure model (PTCM) \citep{Yakovlev}. The MCM explicitly separates cured and susceptible individuals, allowing for independent modelling of the two components, which makes it highly interpretable and flexible. PTCM, on the other hand, avoids explicit classification and models the cure fraction through a hazard-based approach, offering a smoother representation of the transition between risk and cure, but typically requires stronger assumptions on the baseline hazard structure. Given its interpretability and flexibility for modelling, in the following we will focus on the MCM only.

Extending dynamic prediction to settings with a cure fraction has recently attracted growing interest. Most of the existing proposals - with the exception of Shi and Yin (2017)\cite{shi}, who combined landmarking and MCM using the LOCF strategy - have been developed in the context of PTCMs \cite{lambert, beretta, law, kim, barbieri, individual}. The approach of Shi and Yin (2017)\cite{shi} represents, to our knowledge, the only attempt to implement landmarking within an MCM framework. It relies on a simple LOCF-based summary, which may fail to capture the full longitudinal evolution of covariates, especially when the last observation and the landmark time are far apart, and it is unable to correct for the measurement error that is often present when biomarkers are used as covariates. In addition, the adopted definition of a time-dependent cure probability raises conceptual and methodological concerns that will be discussed in more detail in Section \ref{subsec:locf}. In this work, we propose an alternative model-based landmarking method for dynamic prediction in MCMs to address these issues.

The remainder of this paper is organised as follows. In Section~\ref{sec:background}, we introduce the data structure, the notation, and the basics on MCM and landmarking. In Section~\ref{sec:method} we describe the four main steps of the proposed model-based approach. Section~\ref{sec:metrics} describes the metrics used to evaluate predictive performance. In Section~\ref{sec:sim}, we report the design and results of a large-scale simulation study. Section~\ref{sec:case} illustrates the application of the method to a real-world case study. Finally, Section~\ref{sec:discussion} concludes the paper with a discussion of the main findings, limitations, and directions for future research.


\section{Notation and background}\label{sec:background}

\subsection{Data structure and notation}
We consider a dataset with $m$ individuals. For subject $i = 1, \dots, m$, let $\tilde{t}_i$ and $c_i$ denote the true event and censoring times, respectively, and let $t_i = \min(\tilde{t}_i, c_i)$ be the observed time with event indicator $\delta_i = 1$ if $t_i = \tilde{t}_i$ and $\delta_i = 0$ otherwise. 

In addition, we assume that we have observed $p$ covariates (indexed by $l=1,\dots,p$) of two types: 
\begin{itemize}
    \item $p_Y$ \emph{longitudinal covariates} $\bY_i$, measured at multiple time points $t_{ij}$, $i=1,\dots,m$, $j = 1, \dots, n_i$. We denote by $n = \sum_{i=1}^m n_i$ the total number of observations. We allow for a completely unbalanced study design in continuous time, whereby each subject can have a different number of repeated measurements and the measurement times can differ across subjects.
    The vector with the $n_i$ observations of the $l$-th longitudinal covariate for subject $i$ will be denoted by
    \[
    \by_{il} = \big(y_{il}(t_{i1}), \dots, y_{il}(t_{in_{i}})\big)^{\prime}.
    \]
    \item $p-p_Y$ \emph{time-fixed covariates}, which we partition into $\bX_i$ and $\bZ_i$, representing two sets of baseline covariates (possibly overlapping) that will be linked to different components of the MCM.
\end{itemize}

\subsection{Mixture cure models without longitudinal covariates}
MCMs assume that the population is divided into \emph{cured} ($G=0$) and \emph{uncured} ($G=1$) individuals. Cured individuals will never experience the event of interest (e.g. relapse after treatment), whereas uncured individuals may eventually do so. 
However, the cure status $G$ is not always observable in practice, and for each individual the model provides an estimate for the probability of being uncured.

This probability, i.e. the \textit{incidence} component, is described by a binary response model, typically a logistic regression:
\begin{equation}
\label{eqn:logistic} 
P(G_{i}=1 \mid \bX_{i}) = \pi({\bX_{i}}) = \frac{\exp\left(\alpha_0 + \balpha^{\prime} \bX_{i}\right)}{ 1 + \exp\left(\alpha_0 + \balpha^{\prime} \bX_{i}\right)}.
\end{equation}  
For uncured individuals, the survival distribution conditional on covariates $\bZ_i$ is referred to as the \textit{latency} component, and it is commonly specified through a Cox proportional hazards (PH) model:
\begin{equation} 
\label{eqn:uncured} 
S_{u}(t \mid \bZ_{i}) = \exp\left[ -H_{0}(t) \exp\left( \bbeta^{\prime} \bZ_{i} \right) \right],
\end{equation} 
where the subscript $u$ refers to the \emph{uncured} portion of the population and $H_{0}(t)$ is the baseline cumulative hazard. The overall survival function is then 
\begin{equation} 
\label{eqn:overall}
S(t \mid \bX_{i}, \bZ_{i}) = \left[ 1 - \pi(\bX_{i}) \right] + \pi(\bX_{i}) S_u(t \mid \bZ_{i}). 
\end{equation} 
Since this model depends on the latent cure status $G$, the set of parameters is usually estimated using an Expectation-Maximization (EM) algorithm\cite{peng2000nonparametric, sy2000estimation}. \\
Although mixture cure models are not limited to a PH formulation (see Amico and Van Keilegom (2018)\cite{amico} for a review), the Cox PH cure model is a common choice.

\subsection{Landmarking in mixture cure models: the LOCF approach}
\label{subsec:locf}
We consider a mixture cure setting in which landmarking is used to incorporate longitudinal information observed up to a pre-specified time $t_{\ell_{i}}$, referred to as the \emph{landmark time}, to predict survival at a future horizon $t_{\omega_{i}}$. 
The notation $t_{\ell_i}$ allows for subject-specific landmark times, as in settings with a natural landmark (e.g., the time of transplantation for patients on a waiting list). 
In the present work, however, we focus on the simpler case of a common landmark defined in patient time, so that $t_{\ell_i} = t_\ell$ for all individuals.

The only previous attempt to combine landmarking with MCMs is due to Shi and Yin (2017)\cite{shi}, who proposed a class of landmark cure rate models that incorporate both a cure fraction and time-dependent covariates for dynamic prediction. In their approach, the cure probability is specified through a logistic function depending on time-dependent covariates,
\begin{equation}
\label{eqn:shi}    
P(G = 1 \mid Y(t)) = \pi(\bY(t)) = \frac{\exp(\balpha_{0} + \balpha^{\prime}\bY(t))}{1 + \exp(\balpha_{0} + \balpha^{\prime}\bY(t))},
\end{equation}
where the incidence is defined as the limit
\[
P(G = 1) = \lim_{t \to \infty} \pi(\bY(t)).
\]
At a given landmark time $t_{\ell}$, they propose fitting a standard mixture cure model with covariates $\bY(t_{\ell})$.  

While innovative, this construction raises two conceptual issues:

(i) The cure probability is linked to the limiting value $\bY(\infty)$ of time-dependent covariates. This implies that the cure status of an individual depends on future covariate values, even in scenarios where the subject could already have experienced the event before those covariates are realised (e.g., death before transplantation in their motivating example). This specification is not coherent with the idea that the prediction at $t_{\ell}$ should be based only on the information available up to that time.  

(ii) By construction, Shi and Yin (2017)\cite{shi} write the cure probability as in (\ref{eqn:shi}). However, from a probabilistic point of view, it can be shown that
\[
\lim_{t \rightarrow \infty}P(G = 1 \mid \bY(t)) \neq P(G = 1 \mid \bY(t_{\ell}))\]
and the error associated to such an approximation cannot be controlled. Therefore, their specification may lead to theoretical inconsistency.  

To address these limitations, we assume that cure status is determined solely by baseline covariates, while longitudinal covariates influence only the survival distribution of uncured individuals. Specifically, in our case, equations (\ref{eqn:uncured}) and (\ref{eqn:overall}) can be rewritten, respectively, as
\begin{equation} 
\label{eqn:uncuredbis} 
S_{u}(t \mid \bZ_{i}, \bY_{i}) = \exp\left[ -H_{0}(t) \exp\left( \bbeta^{\prime} \bZ_{i} + \bpsi^{\prime} \bY_{i}(t) \right) \right], \end{equation}
and
\begin{equation} 
S(t \mid \bX_{i}, \bZ_{i}, \bY_{i}) = \left[ 1 - \pi(\bX_{i}) \right] + \pi(\bX_{i}) S_u(t \mid \bZ_{i}, \bY_{i}(t)). 
\end{equation} 
This ensures that the prediction at the landmark time relies exclusively on the information available up to that point. Furthermore, the cure status is constant over time, and the corresponding probability is defined in a coherent and identifiable way.
In fact, it is often clinically reasonable to regard the propensity to cure as fixed at baseline, reflecting inherent patient- or disease-specific characteristics rather than post-baseline evolution.

\section{Model-based landmarking in cure models}\label{sec:method}

We propose a model-based landmarking method for MCMs in which a statistical model is used to describe the dynamics of longitudinal covariates up to the landmark time. The summaries derived from these models are then combined with the baseline information for prediction. The procedure follows four main steps, inspired by Signorelli et al. (2021)\cite{pencal}:
\begin{enumerate}
    \item model the longitudinal covariates via (G)LMMs;
    \item summarize the longitudinal histories through predicted random effects;
    \item estimate the MCM using an EM algorithm;
    \item generating post-landmark predictions.
\end{enumerate}
We describe these steps in detail below.

\subsection{Step 1: modelling the longitudinal covariates}
The dynamics in the longitudinal covariates up to the landmark time are described by using either a linear mixed model (LMM) \citep{lmm} or a generalized linear mixed-effects model (GLMM) \citep{glmm}.

For each longitudinal covariate, we fit a (generalized) LMM. For non-Gaussian responses, we follow the formulation of Achana \emph{et al.}\citep{mglmm}, which linearizes the model regardless of the response and provides an alternative implementation of penalized quasi-likelihood in the spirit of Breslow and Clayton\citep{pql1993}. The model is fitted using only measurements collected before the landmark time $t_{\ell_{i}}$ and restricted to individuals still at risk at $t_{\ell_{i}}$:
\begin{equation*}
Y_{il}(t_{ij}) = g^{-1} \biggl( \bW_{il}(t_{ij}) \bgamma_l + \bV_{il}(t_{ij}) \bb_{il} \biggr) + \epsilon_{il}(t_{ij}),
\quad t_{ij} \leq t_{\ell_{i}},
\end{equation*}
$i=1,\dots,m$, $j=1,\dots,n_{i}(t_{\ell_{i}})$, $l=1,\dots,p_{Y}$, where $n_{i}(t_{\ell_{i}})$ denotes the number of measurements available for subject $i$ up to $t_{\ell_{i}}$. Here
\begin{itemize}
\item $g(\cdot)$ is a differentiable monotonic link function;
\item $\bgamma$ is a vector of fixed effects;
\item $\bb_{il} \sim MVN(0,\bSigma_{b_l})$ is a vector of individual-specific random effects, capturing heterogeneity across subjects;
\item $\bW$ is a design matrix with size $m \times p_W$; it usually contains elements of $\bX, \bZ$ and polynomials in $t$;
\item $\bV \subseteq \bW$ is a design matrix with size $m \times p_{V}$. Tipically it includes a constant term (corresponding to an individual-specific intercept) and polynomials in $t$ to describe individual-specific dynamics, as in latent growth models \citep{bollen};
\item $\bepsilon_{il} \sim MVN(0, \bSigma_{\epsilon_{il}})$ is a Gaussian error term, with $\bSigma_{\epsilon_{il}}$ typically specified as $\sigma_{l}^{2}\bI_{n_{i}(t_{\ell_{i}})}$, assumed to be conditionally independent of $\bb_{il}$ and uncorrelated with $\bW_{il}$.
\end{itemize}

\subsection{Step 2: summarizing longitudinal covariates at the landmark time}
At the landmark time, each patient’s longitudinal history is summarized using the predicted random effects $\hat{\bb}_{il}$ obtained from the (G)LMM fitted up to $t_{\ell_{i}}$. These random effects provide a concise representation of individual-specific heterogeneity in observed dynamics and can be treated as time-constant predictors in the cure model. 

For the LMM case, the predicted random effects $\hat{\bb}_{il}$ are calculated as follows:
\begin{equation}
    \hat{\bb}_{il} = \mathbb{E}\ (\bb_{il} \mid \by_{il}) = \hat{\bSigma}_{b_l} \bV_{il}^{\prime}\left[\bV_{il} \hat{\bSigma}_{b_l} \bV_{il}^{\prime} + \hat{\bSigma}_{\epsilon_{il}} \right]^{-1} (\by_{il} - \bW_{il} \hat{\bgamma}_l)
\end{equation}
that is, the estimator derived using the conditional distribution of $\bb_{il} \mid \by_{il}$ obtained exploiting the joint Gaussianity of $(\by_{il},\bb_{il})$. 
The variance components $\hat{\bSigma}_{b_l}$ and $\hat{\bSigma}_{\epsilon_{il}}$ are estimated from the fitted LMM via restricted maximum likelihood (REML).

When the longitudinal outcome does not satisfy the Gaussian assumption, we fit a generalized linear mixed model. \\ 
The marginal density of $\by_{il}$ is given by
\begin{equation}
\label{eqn:marginal}
    f_{Y}(\by_{il}) = \int f_{Y}(\by_{il} \mid \bb_{il}) f_{B}(\bb_{il}) \, d\bb_{il},
\end{equation}
where $f_{Y}(\by_{il} \mid \bb_{il})$ is the conditional density specified by the regression model as a function of covariates $\bW_{il}$ and random effects $\bb_{il}$, and $f_{B}(\bb_{il})$ is the prior density for random effects.  
In both LMMs and GLMMs, if the correlation between observed covariates $\bW_{il}$ and unobserved heterogeneity $\bb_{il}$ is an issue, it can be addressed through a Mundlak-type reparameterization \citep{mundlak}, adding cluster-level means of time-varying covariates to the fixed effects part of the model.

Since the integral in the marginal density (\ref{eqn:marginal}) does not have a closed form, a numerical approximation is required. Following Wolfinger and O'Connell\cite{pseudo}, we adopt a \emph{penalized quasi-likelihood} (PQL) approach, which relies on a first-order Taylor expansion to linearize the model on the link scale. Specifically, we define the linearized response
\begin{equation}
\by_{il}^{*} = \bW_{il} \bgamma_{l} + \bV_{il}\bb_{il} + \bepsilon_{il}^{*},
\end{equation}
where $\by_{il}^*$ approximates $\by_{il}$ on the link scale for the $l$-th longitudinal covariate, and
\begin{equation}
\bepsilon_{il}^{*} = [g^{'}(\hat{\bmu}_{il})]^{-1}(\by_{il} - \hat{\bmu}_{il}),
\end{equation}
with $\hat{\bmu}_{il}$ denoting the fitted mean.  
After linearization, $\bepsilon_{il}^*$ is assumed to follow a multivariate Gaussian distribution with zero mean and covariance matrix $\bSigma_{\bepsilon_{il}^{*}}$, parametrized analogously to $\bSigma_{\epsilon_{il}}$ in the LMM. The resulting conditional and marginal distributions of $\by_{il}^*$ are
\begin{equation}
\by_{il}^{*} \mid \bb_{il} \sim MVN(\bW_{il} \bgamma_{l} + \bV_{il}\bb_{il}, \bSigma_{\bepsilon_{il}^{*}})
\end{equation}
and
\begin{equation}
\by_{il}^{*} \sim MVN(\bW_{il} \bgamma_{l}, \bV_{il} \bSigma_{b_{l}} \bV_{il}^{\prime} + \bSigma_{\bepsilon_{il}^{*}}).
\end{equation}
As in the LMM case, the predicted random effects $\hat{\bb}_{il}$ are then obtained as empirical Bayes estimates from the conditional distribution of $\bb_{il} \mid \by_{il}^*$.

\subsection{Step 3: estimation of the cure model parameters}
Once summaries for observed time-dependent covariates have been obtained, we consider the set of subjects still at risk at the landmark time and employ a mixture cure model for post-landmark survival prediction, incorporating both baseline covariates $(X_i, Z_i)$ and summaries $\hat{\textbf{b}}_i = (\hat{b}_{i1}, \dots, \hat{b}_{ip_Y})$. Specifically, while the incidence component is as in (\ref{eqn:logistic}), the latency in (\ref{eqn:uncured}) becomes
\begin{equation*}
    S_u(t \mid Z_i, \hat{b}_i) = \exp [-H_0(t) \exp(\bbeta^{\prime}Z_i + \bpsi^{\prime}\hat{b}_i)].
\end{equation*}

If we denote by $\btheta$ the \emph{complete} set of parameters of interest, the complete-data likelihood function for the Cox PH cure model can be written as 
\begin{align}
\label{eqn:like}
   \mathcal{L}_{c}(\btheta)=\prod_{i=1}^{n} &
[\pi(\bX_i) f_u(T_i \mid \bZ_i, \bb_i)] ^ {\delta_i G_i} 
[\pi(\bX_i) S_u(T_i \mid \bZ_i, \bb_i)] ^ {(1-\delta_i) G_i} \notag \\
& [1-\pi(\bX_i)] ^ {(1-\delta_i) (1-G_i)}
\end{align} 
In the E-step of the EM algorithm\citep{em}, we calculate the posterior expectation of the latent indicator $G_i$. For individual $i \in (1,...m)$ at iteration $r$, this is given by
\begin{align}
\label{estep}
q_i^{(r)} = \delta_i+(1- \delta_i)\frac{\pi^{(r-1)}(\bX_i) S_u^{(r-1)}(T_i \mid \bZ_i, \hat{\bb}_i)} {\left[1-\pi^{(r-1)}(\bX_i)\right]+\pi^{(r-1)}(\bX_i) S_u^{(r-1)}(T_i \mid \bZ_i, \hat{\bb}_i)}
\end{align}
The M-step maximizes the expected log-likelihood replacing $G_i$ by $q_i$ and optimizing over the parameters set. 

The \emph{loss} function used in the M-step is the negative of the conditional expectation of the complete data log-likelihood given the observed data and the current parameter estimates $\btheta^{(r)}$. This loss function can be decomposed into the sum of two loss sub-functions, one for the incidence and the other for the latency components. 
That is, 
\[
\mathbb{E}\ \left\{\log \left[\mathcal{L}_{c}(\cdot) \mid \text{data}, \btheta^{(r)} \right]\right\} =Q_{1}\left(\btheta \mid \btheta^{(r)}\right) + Q_{2}\left(\btheta \mid \btheta^{(r)}\right)
\]
where the expected log-likelihood contribution for the incidence component can be written as 
\begin{equation}
Q_{1}\left(\btheta \mid \btheta^{(r)}\right)= \sum_{i=1}^{n} q_i^{(r)} \log \pi (\bX_i) + (1-q_i^{(r)})(1 - \delta_i) \log (1 - \pi(X_i)) 
\end{equation}
and $(1-q_i^{(r)})(1 - \delta_i)=(1 - \delta_i)$ by definition. The contribution for the latency component is
\begin{equation}
Q_{2}\left(\btheta \mid \btheta^{(r)}\right) = \sum_{i=1}^{n} q_i^{(r)} \delta_i \log f_u(T_i\mid \bZ_i, \hat{\bb}_i) + q_i^{(r)} (1 - \delta_i) \log S_u (T_i \mid \bZ_i, \hat{\bb}_i)
\end{equation}
Maximization with respect to $H_0$ at iteration $r$ leads to the following modified Breslow estimator \citep{breslow} of the baseline cumulative hazard function 
\begin{align}
\label{eqn:H0}
\hat{H}_0^{(r)}(t)={\sum\limits_{i=1}^{m}} \frac {I(t_{i}<t)\delta_{i}} {{\sum\limits_{i \in R_t}} q_i^{(r)} \exp{(\bbeta^{(r-1)\prime}\bZ_i + \bpsi^{(r-1)\prime}\hat{\bb}_i)}}
\end{align}
where the numerator is the number of events up to time $t$ and $R_t$ represents the set of subjects that are still at risk at time $t$. 

\subsection{Step 4: post-landmark prediction}
Once the regression parameters have been estimated, we can derive the conditional cure and survival probabilities by plugging in the estimated values:
\begin{equation}
\hat{\pi}(\bX_{i} \mid t_{\ell_{i}})=\frac{\exp(\hat{\balpha}_0 + \hat{\balpha}^{\prime} \bX_{i})}{1 + \exp(\hat{\balpha}_0 + \hat{\balpha}^{\prime} \bX_{i})},
\end{equation}
\begin{equation}
\hat{S}{u}(t_{\omega_{i}} \mid \bZ_{i}, \hat{\bb_{i}}, t_{\ell_{i}}) = \exp \biggl(-\hat{H}_0 (t_{\omega_{i}}), \exp (\hat{\bbeta}^{\prime} \bZ_{i} + \hat{\bpsi}^{\prime} \hat{\bb_{i}} )\biggr).
\end{equation}
Predictions can be obtained for both subjects used in model estimation and for out-of-sample individuals, since they rely on the parameter estimates obtained in Step 3. This naturally raises the question of how to evaluate the quality of such predictions. The next section describes the accuracy measures that we suggest employing for this purpose.


\section{Metrics for predictive performance evaluation}\label{sec:metrics}

In the MCM framework, predictive performance is typically assessed separately\cite{asano} for the \emph{incidence} and the \emph{latency} component.

For the incidence component, since the true cure status $G_i$ is only partially observed, we rely on weighted versions of standard classification metrics, where each subject is weighted by its estimated posterior probability of being uncured $\hat{q}_i$. Discrimination is quantified by the \emph{weighted AUC}\cite{roc2},
\begin{equation*}
    \widehat{\mathrm{AUC}}_{inc} = \frac{\sum_{i \neq j} I(\hat{\eta}^{(I)}_{i} > \hat{\eta}^{(I)}_{j})\,\hat{q}_i\,(1-\hat{q}_j)}{\sum_{i\neq j} \hat{q}_i\,(1-\hat{q}_j)},
\end{equation*}
where $\hat{\eta}^{(I)}_{i}=\hat{\alpha}_0 + \hat{\balpha}^{\prime} \bX_{i}$ is the incidence linear predictor and $\hat{q}$ is the posterior expectation in (\ref{estep}) obtained at the convergence of the EM algorithm. Prediction error is measured by the \emph{weighted Brier score}\cite{graf1999},
\begin{equation*}
    \widehat{\mathrm{BS}}_{inc} = \frac{1}{m}\sum_{i = 1}^{m} \hat{u}_i \, (\hat{q}_i - \tilde{G}_i)^2,
\end{equation*}
where $\tilde{G}_i=1$ if the subject $i$ experienced the event, $\tilde{G}_i=0$ if censored beyond the largest observed event time, and $\hat{u}_i$ are the inverse probability of censoring weights (IPCW), i.e. the inverse of the Kaplan–Meier estimate of the probability of remaining uncensored up to the observed time.

For the latency component, prediction performance is evaluated using standard survival measures modified for censoring. Discrimination at time $t$ is assessed by the \emph{time-dependent AUC}\cite{heagerty2000},
\begin{equation*}
    \widehat{\mathrm{AUC}}(t)_{lat} = \frac{\sum_{i \neq j} I(\hat{\eta}^{(L)}_{i} > \hat{\eta}^{(L)}_{j})\,\delta_i(t)\,(1-\delta_j(t))\,\hat{u}_{i}(t)\,\hat{u}_{j}(t)}{\sum_{i\neq j} \delta_i(t)\,(1-\delta_j(t))\,\hat{u}_{i}(t)\,\hat{u}_{j}(t)},
\end{equation*}
where $\hat{\eta}^{(L)}_{i} = \hat{\bbeta}^{\prime} \bZ_{i} + \hat{\bpsi}^{\prime} \hat{\bb}_{i}$ is the latency linear predictor and $\delta_i(t)$ indicates whether subject $i$ experienced the event before $t$. Prediction error at time $t$ is measured by the \emph{time-dependent Brier score}\cite{schoop},
\begin{equation*}
    \widehat{\mathrm{BS}}(t)_{lat} = \frac{1}{m} \sum_{i=1}^m \hat{u}_i(t) \, \left( I(t_i>t) - \hat{S}_i(t) \right)^2,
\end{equation*}
where $\hat{S}_i(t)$ is the predicted survival probability for subject $i$ and $\hat{u}_i(t)$ are the corresponding IPCW weights. Finally, the overall discriminative ability can be summarized by the \emph{concordance index (C-index)}\cite{harrell1982, uno2011},
\begin{equation*}
    \widehat{C} = \frac{\sum_{i \neq j} I(\hat{\eta}^{(L)}_{i} > \hat{\eta}^{(L)}_{j})\,D_{i,j}}{\sum_{i\neq j} D_{i,j}},
\end{equation*}
with $D_{i,j} = I[t_i < t_j,\,\delta_i = 1] + I[t_i = t_j,\,\delta_i=1,\,\delta_j=0]$.


\section{Simulation study}\label{sec:sim}
\subsection{Simulation design}\label{sec:design}
We conducted a simulation study to evaluate the performance of the proposed model-based approach for dynamic prediction in mixture cure models, comparing it with a LOCF strategy. The study considered different sample sizes, cure fractions, number of repeated measurements, and data-generating mechanisms.

For each setting, 1000 datasets including four time-invariant and four time-varying covariates were generated. Subjects were followed longitudinally up to a pre-specified landmark time ($t_\ell = 3$), with $M_i$ repeated measurements of the time-varying covariates. The time-varying covariates were simulated as linear trajectories with subject-specific random intercepts and slopes drawn from normal distributions, $N(0,1)$ and $N(0,0.7)$ respectively, and standard Gaussian measurement-level noise was added, $\varepsilon_{il} \sim N(0,1)$. Two longitudinal designs were considered: a \textit{balanced} design with $M_i = 10$ measurements per subject, and an \textit{unbalanced} design obtained by randomly deleting some of the observation times so that $M_i \in [5,10]$. Under the unbalanced design, the LOCF strategy, applied after the amputation step, relied on the last observed measurement after deletion, which could occur well before the landmark. From the landmark onwards, all subjects were considered at risk, with event times simulated from a Weibull distribution conditional on the time-varying covariates and regression coefficients $\boldsymbol{\beta} = (1, 0, -1, 0)$. Independent censoring times were drawn from an exponential distribution starting at the landmark.

To assess the robustness of the proposed model to misspecification, three data-generating mechanisms were considered for the latency component: (i) a \textit{strongly misspecified} setting, where time-varying covariates and survival times were linked using $\bY(t)$; (ii) a \textit{mildly misspecified} setting, where the hazard model in the latency component depends on the current value of the longitudinal process $\bY(t_{\ell})$; and (iii) a \textit{correctly specified model}, where event times are generated using the random effects approach underlying the proposed method. For each of the 1000 simulated datasets, predictions were obtained on an external validation dataset, generated independently but under the same data-generating mechanism as the corresponding training dataset.

Time-invariant covariates are simulated from a multivariate normal distribution, $N_4(\boldsymbol{0}, \mathbf{I})$. The cure status was drawn from a Bernoulli distribution via a logistic model for the probability of \emph{not} being cured, conditional on time-invariant covariates. To vary the cure fraction, only the intercept was modified, while the remaining coefficients were fixed at $(-1, 0, 1, 0)$. Specifically, an intercept value $\alpha_{0} = 2$ yielded, on average, a cure fraction of 20\%, whereas $\alpha_{0} = 0.65$ yielded 40\%.

Each combination of cure fraction and longitudinal design (balanced or unbalanced) was tested at two sample sizes ($m = 300,\ 1000$), resulting in four basic scenarios replicated twice.

An overview of all simulation scenarios is provided in Table~\ref{tab:sim_scenarios}. All code and materials to reproduce the simulation study are available in a dedicated GitHub repository at \url{https://github.com/martacip/cure_prediction}.

\subsection{Simulation results}
The results for the incidence component are reported in Table~\ref{tab:incidence_results_n300_n1000}. Since both LOCF and the model-based approach rely on the same specification for the cure model, their predictive performance in terms of incidence is essentially indistinguishable across scenarios. Small fluctuations are observed, but these appear to be random and unrelated to the sample size or design and do not seem to indicate any systematic advantage of either method.

\begin{figure}[ht]
\centering
\textbf{$m=300$}\\
\subfloat[Scenario 1 - 20\% cured; $M$ balanced]{\includegraphics[width=.4\linewidth]{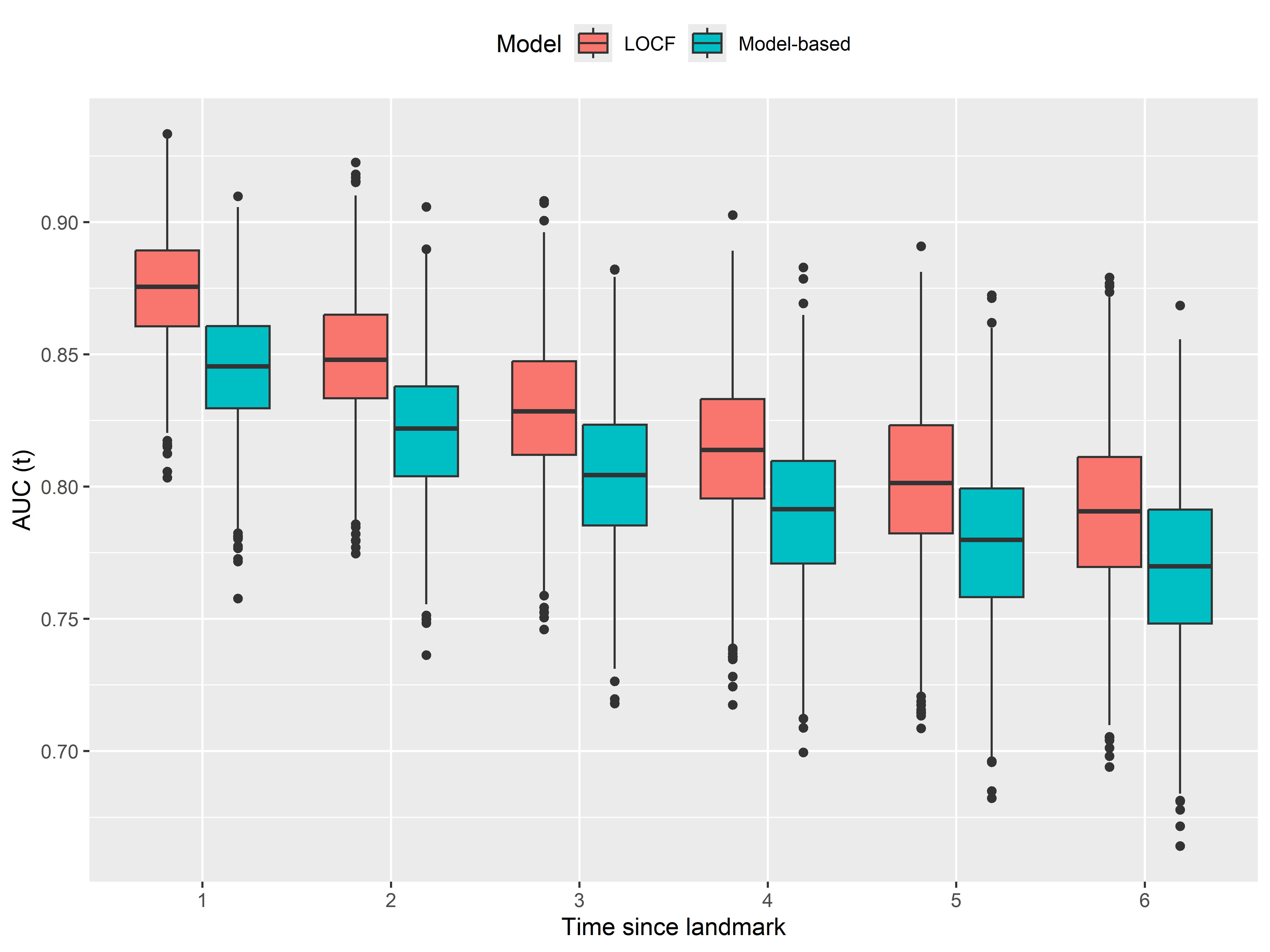}} 
\subfloat[Scenario 2 - 20\% cured; $M$ unbalanced]{\includegraphics[width=.4\linewidth]{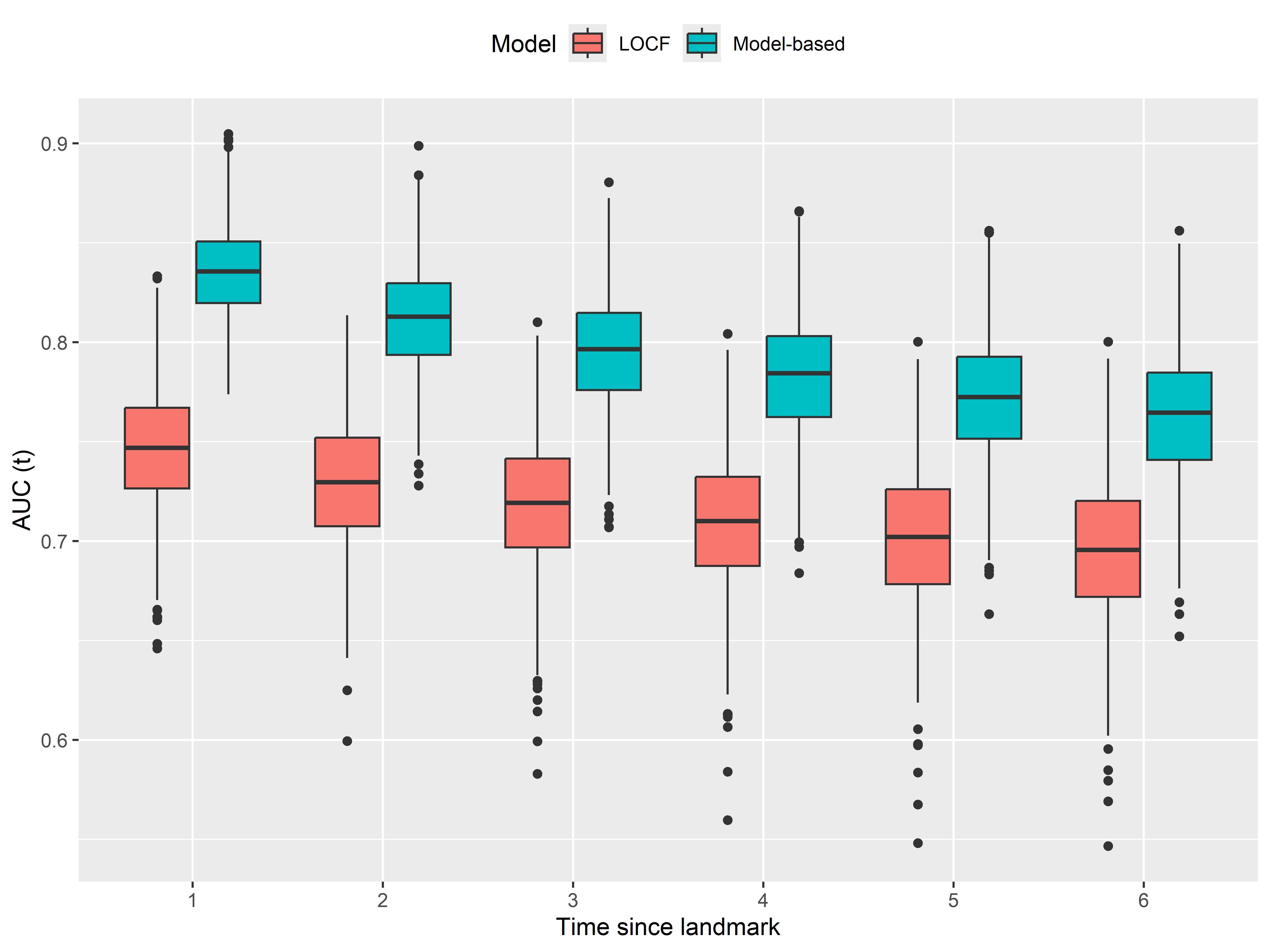}} \\
\subfloat[Scenario 3 - 40\% cured; $M$ balanced]{\includegraphics[width=.4\linewidth]{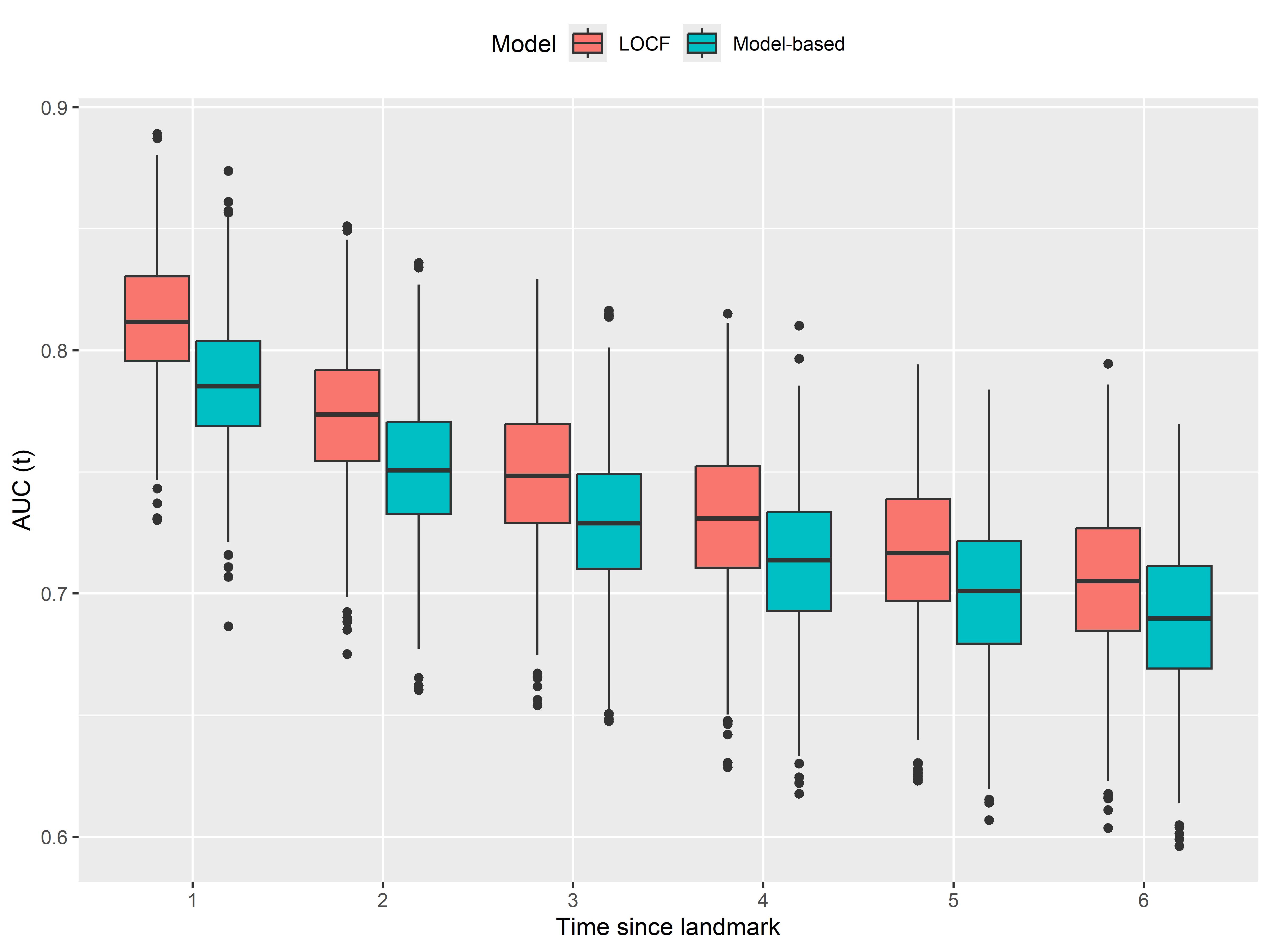}} 
\subfloat[Scenario 4 - 40\% cured; $M$ unbalanced]{\includegraphics[width=.4\linewidth]{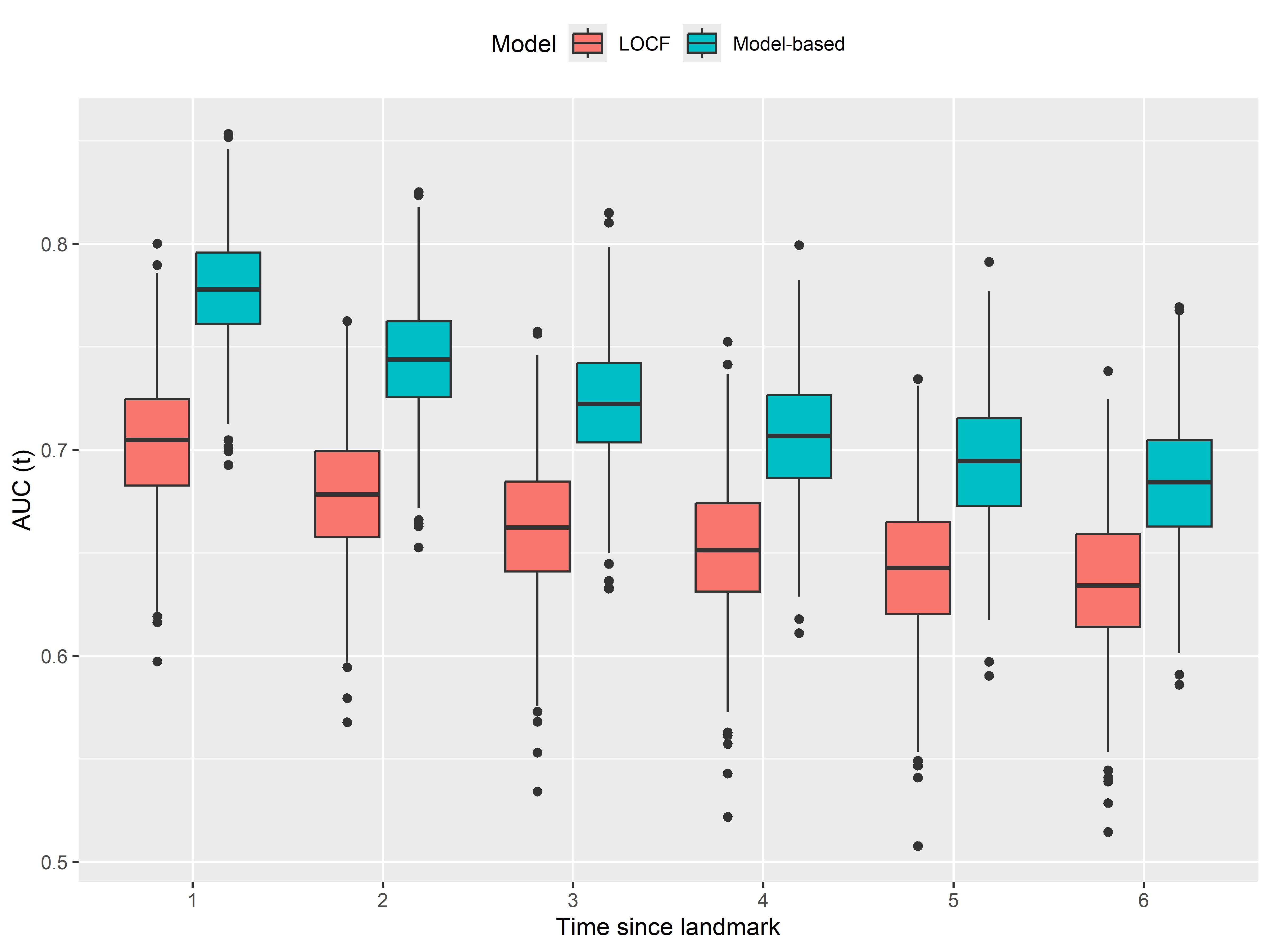}} \\[2mm]
\textbf{$m=1000$}\\
\subfloat[Scenario 1 - 20\% cured; $M$ balanced]{\includegraphics[width=.4\linewidth]{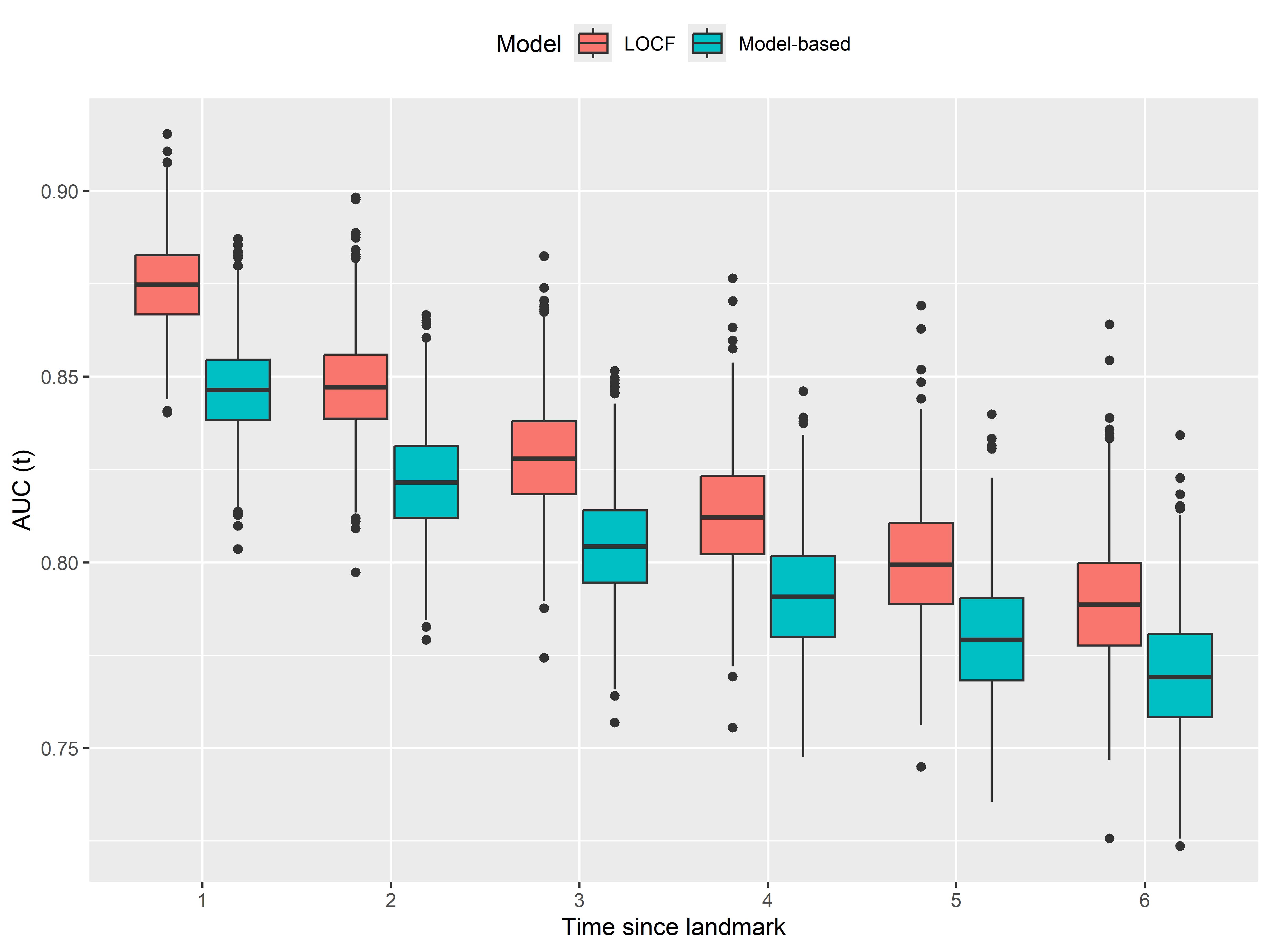}} 
\subfloat[Scenario 2 - 20\% cured; $M$ unbalanced]{\includegraphics[width=.4\linewidth]{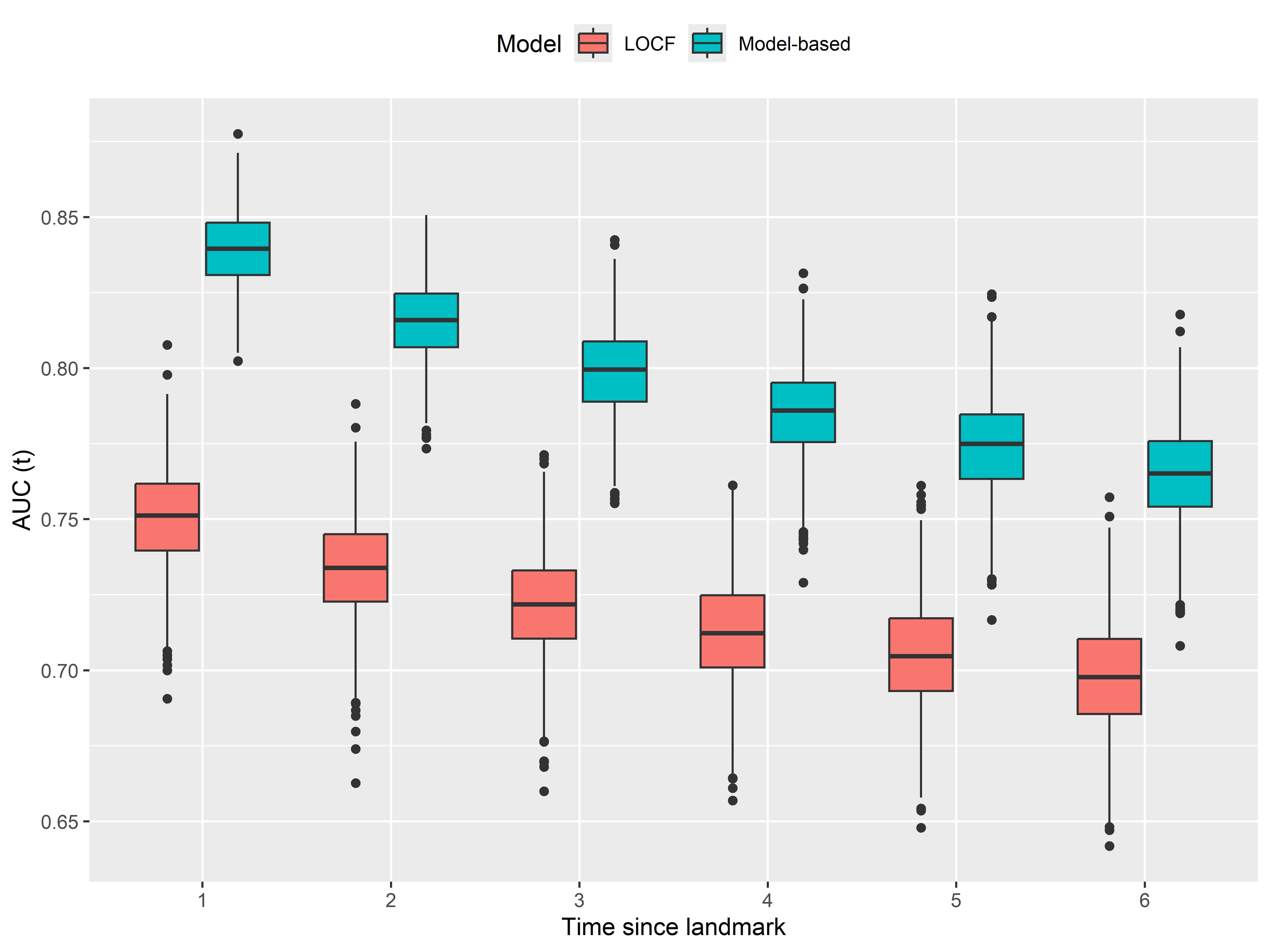}} \\
\subfloat[Scenario 3 - 40\% cured; $M$ balanced]{\includegraphics[width=.4\linewidth]{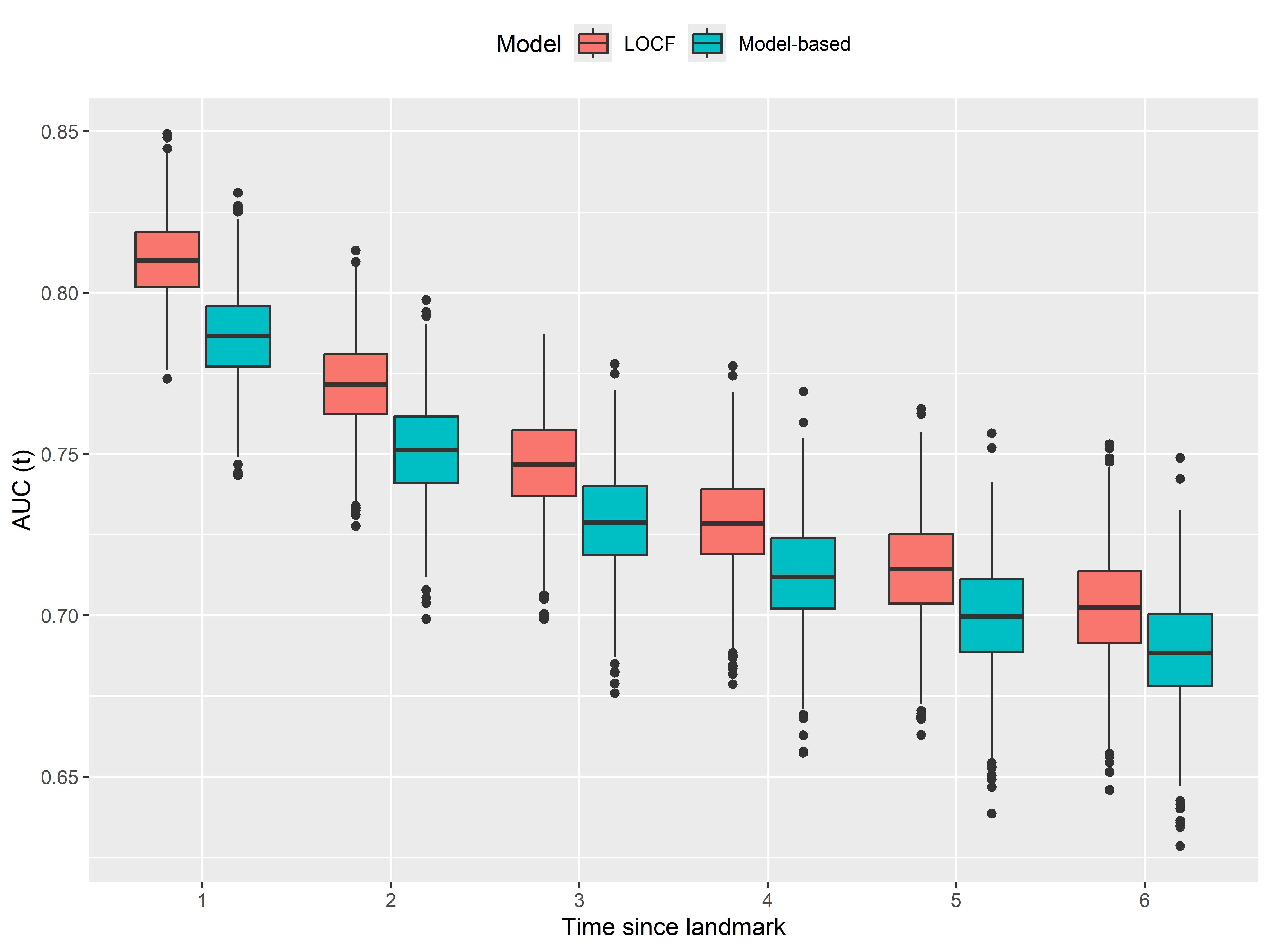}} 
\subfloat[Scenario 4 - 40\% cured; $M$ unbalanced]{\includegraphics[width=.4\linewidth]{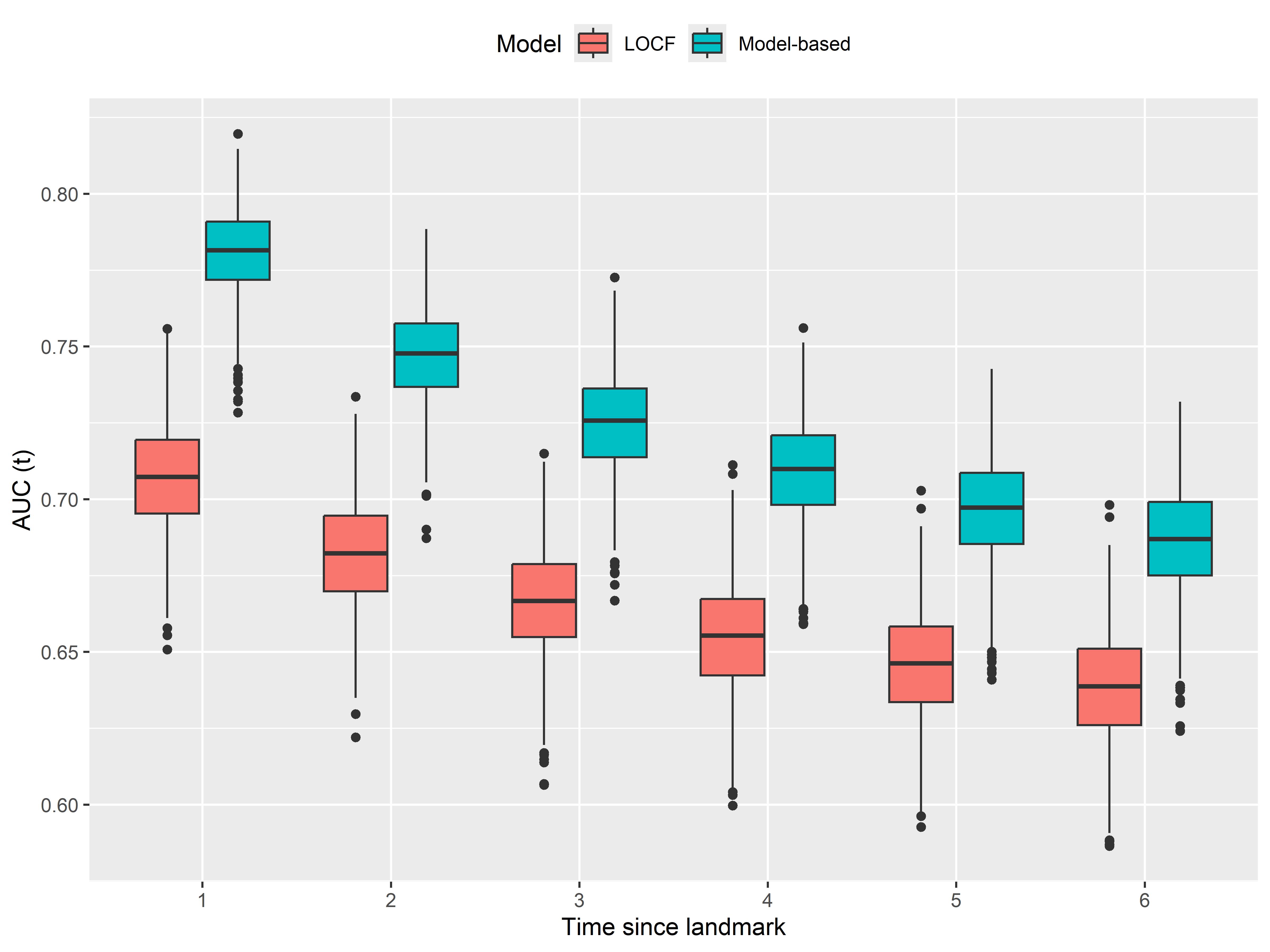}} 
\caption{Time-dependent AUC for scenarios under strong misspecification.}
\label{fig:strong}
\end{figure}

\begin{figure}[ht]
\centering
\textbf{$m=300$}\\
\subfloat[Scenario 5 - 20\% cured; $M$ balanced]{\includegraphics[width=.4\linewidth]{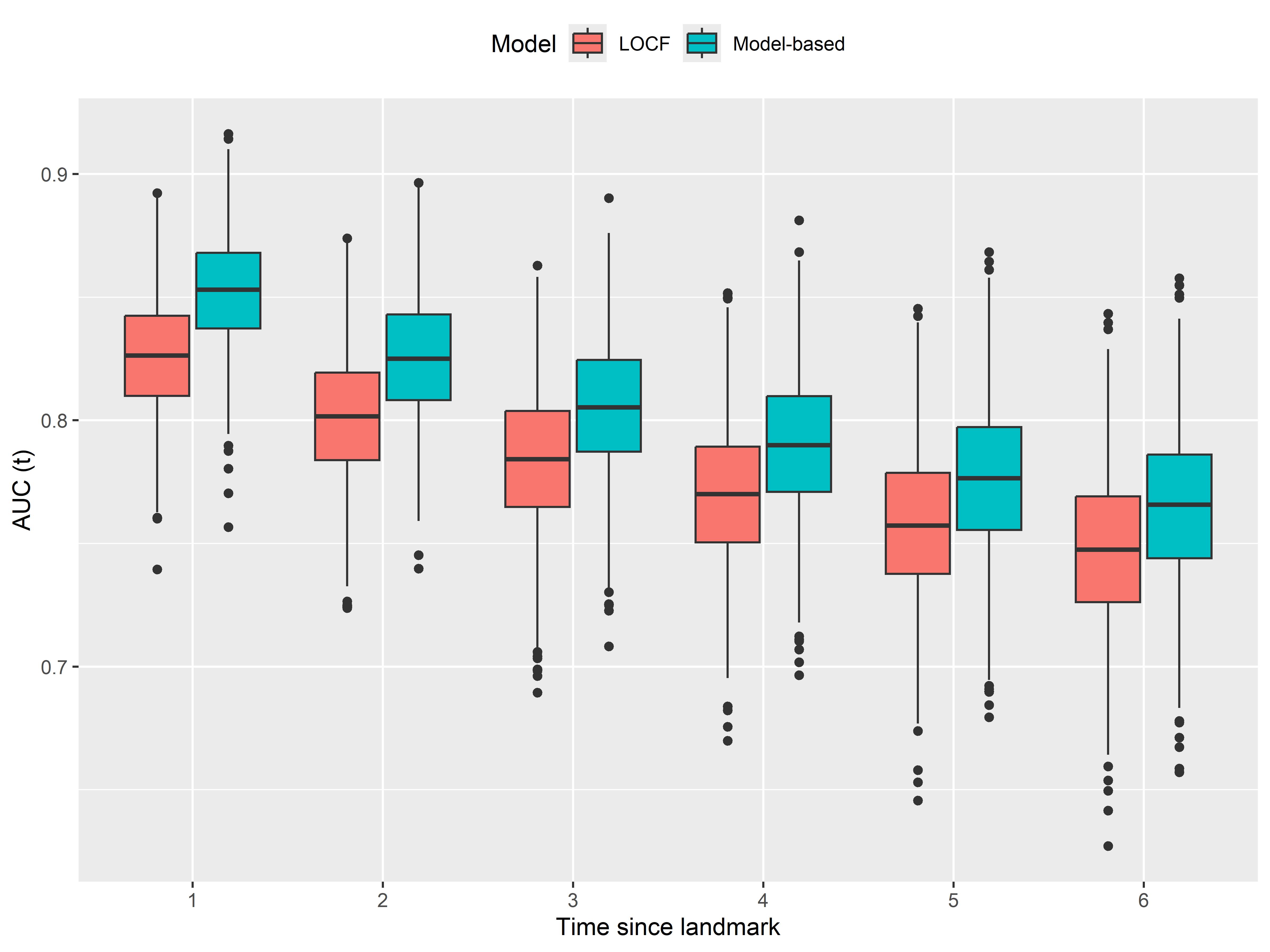}} 
\subfloat[Scenario 6 - 20\% cured; $M$ unbalanced]{\includegraphics[width=.4\linewidth]{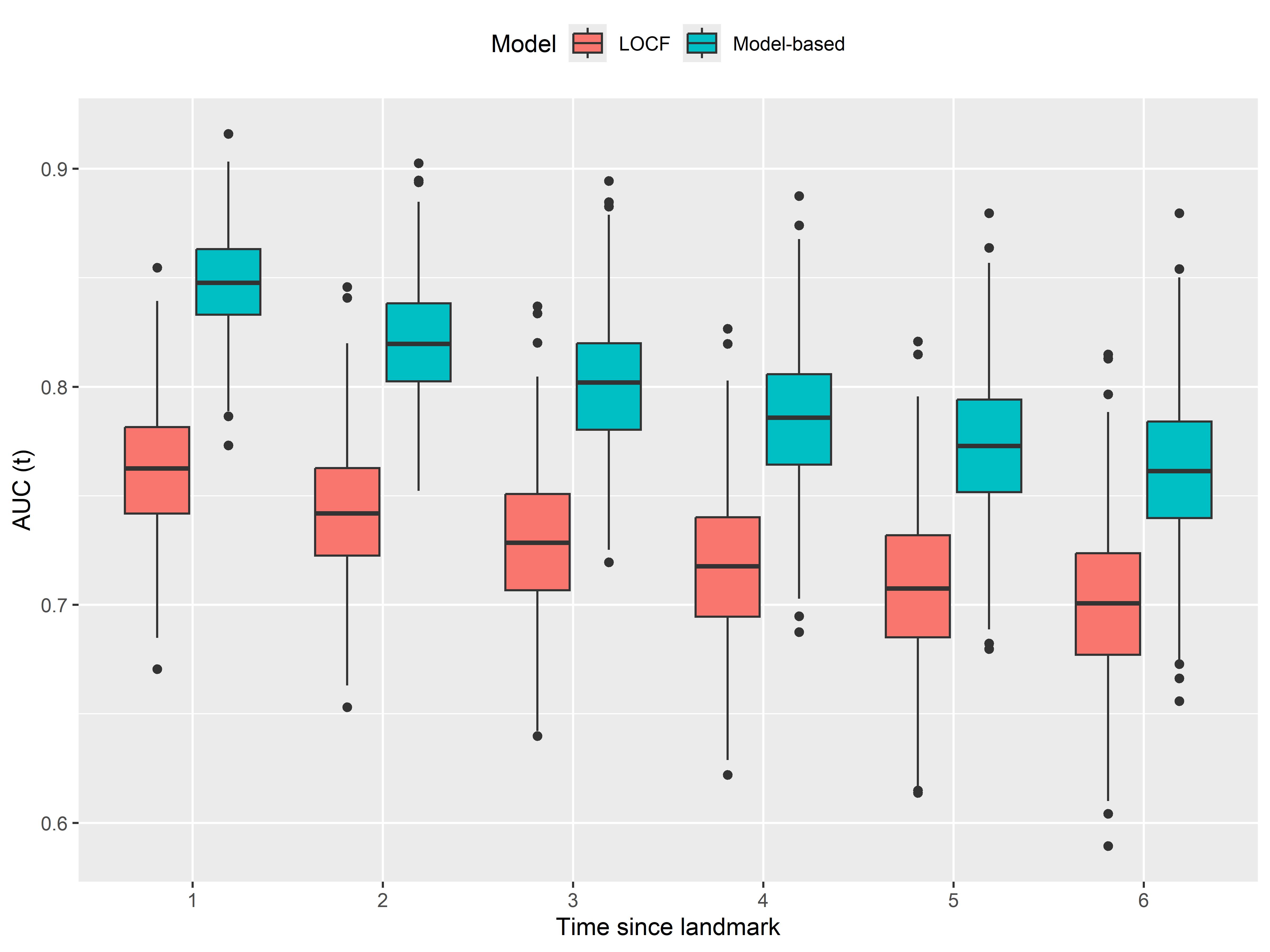}} \\
\subfloat[Scenario 7 - 40\% cured; $M$ balanced]{\includegraphics[width=.4\linewidth]{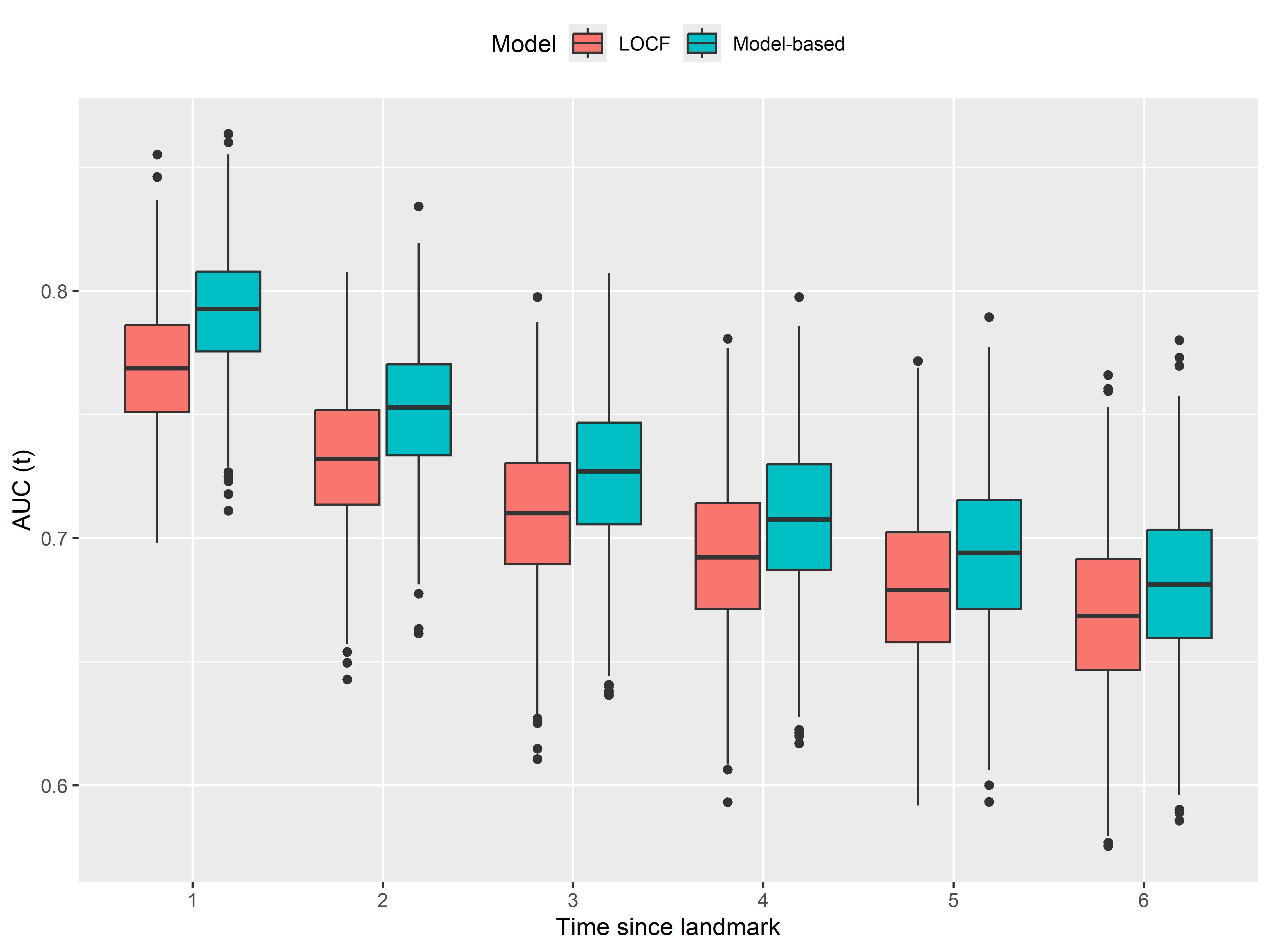}} 
\subfloat[Scenario 8 - 40\% cured; $M$ unbalanced]{\includegraphics[width=.4\linewidth]{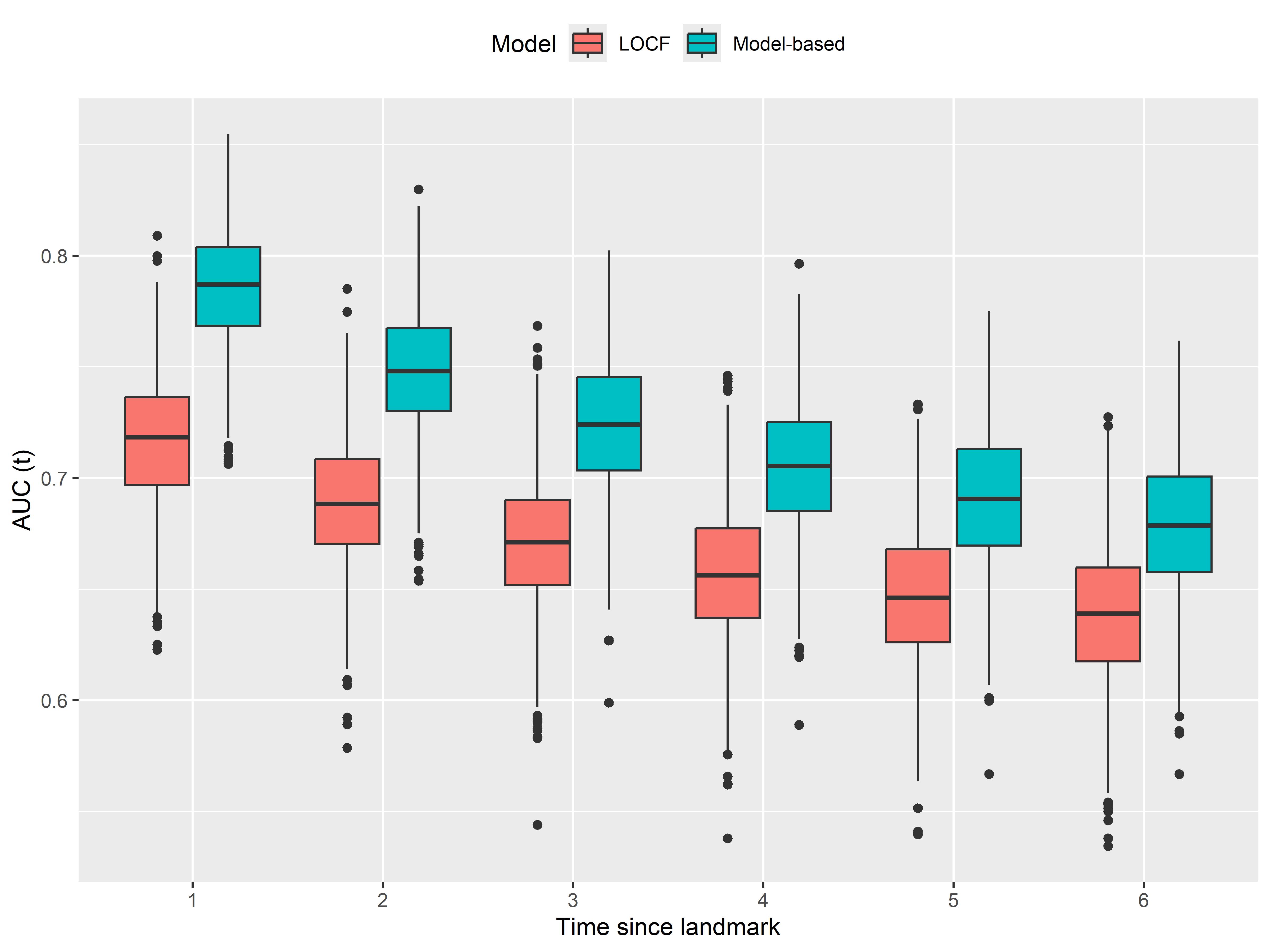}} \\[2mm]
\textbf{$m=1000$}\\
\subfloat[Scenario 5 - 20\% cured; $M$ balanced]{\includegraphics[width=.4\linewidth]{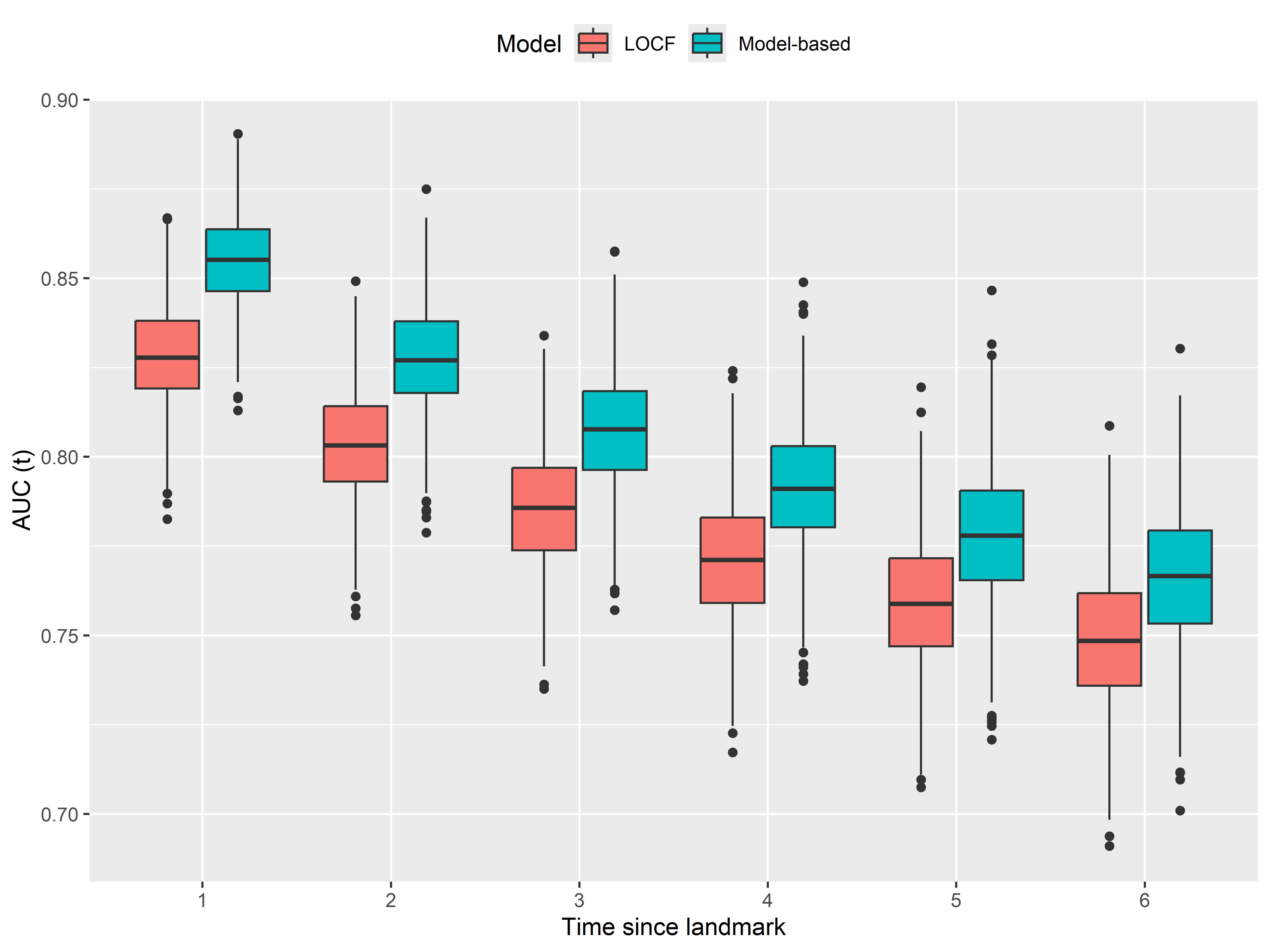}} 
\subfloat[Scenario 6 - 20\% cured; $M$ unbalanced]{\includegraphics[width=.4\linewidth]{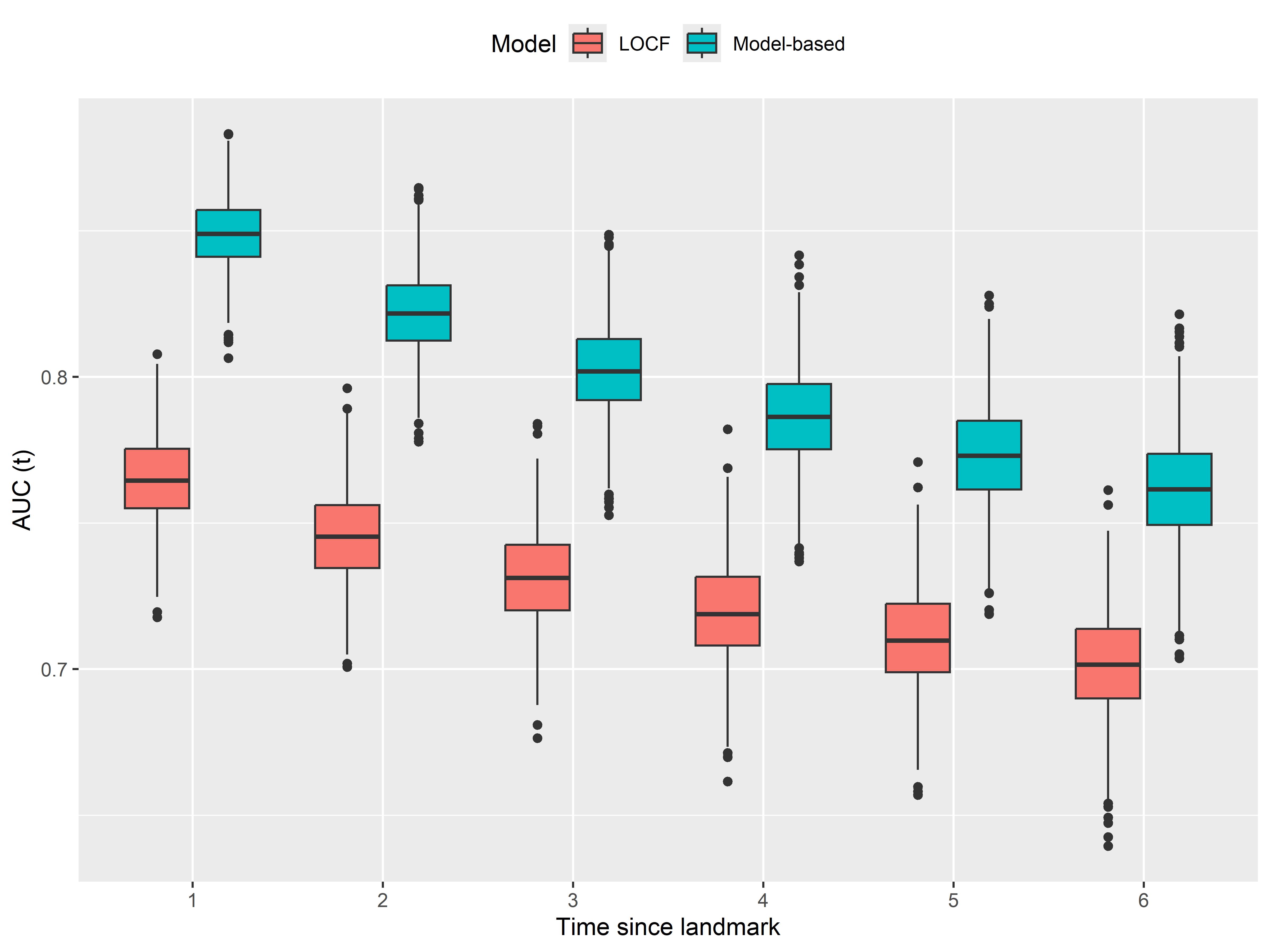}} \\
\subfloat[Scenario 7 - 40\% cured; $M$ balanced]{\includegraphics[width=.4\linewidth]{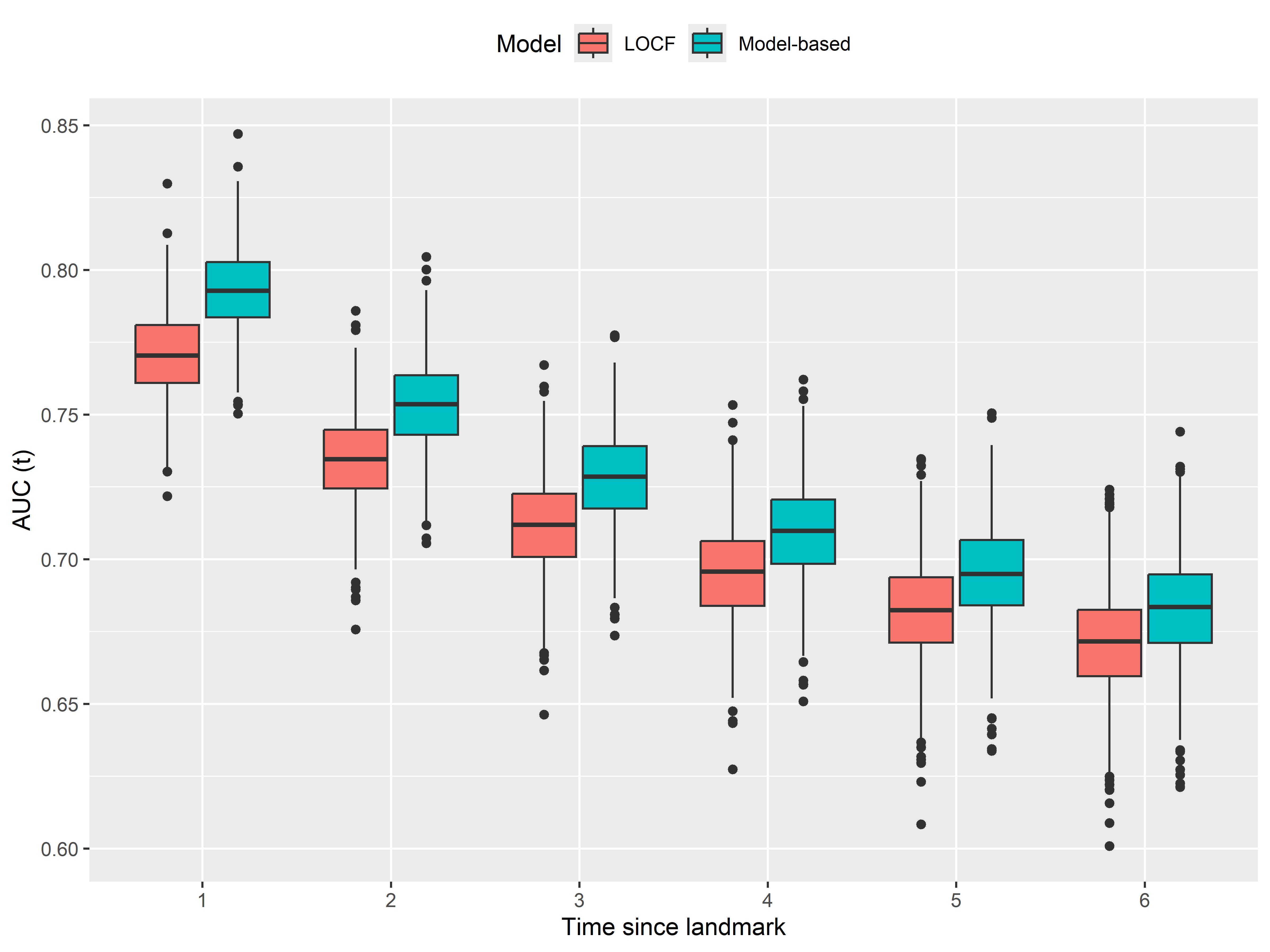}} 
\subfloat[Scenario 8 - 40\% cured; $M$ unbalanced]{\includegraphics[width=.4\linewidth]{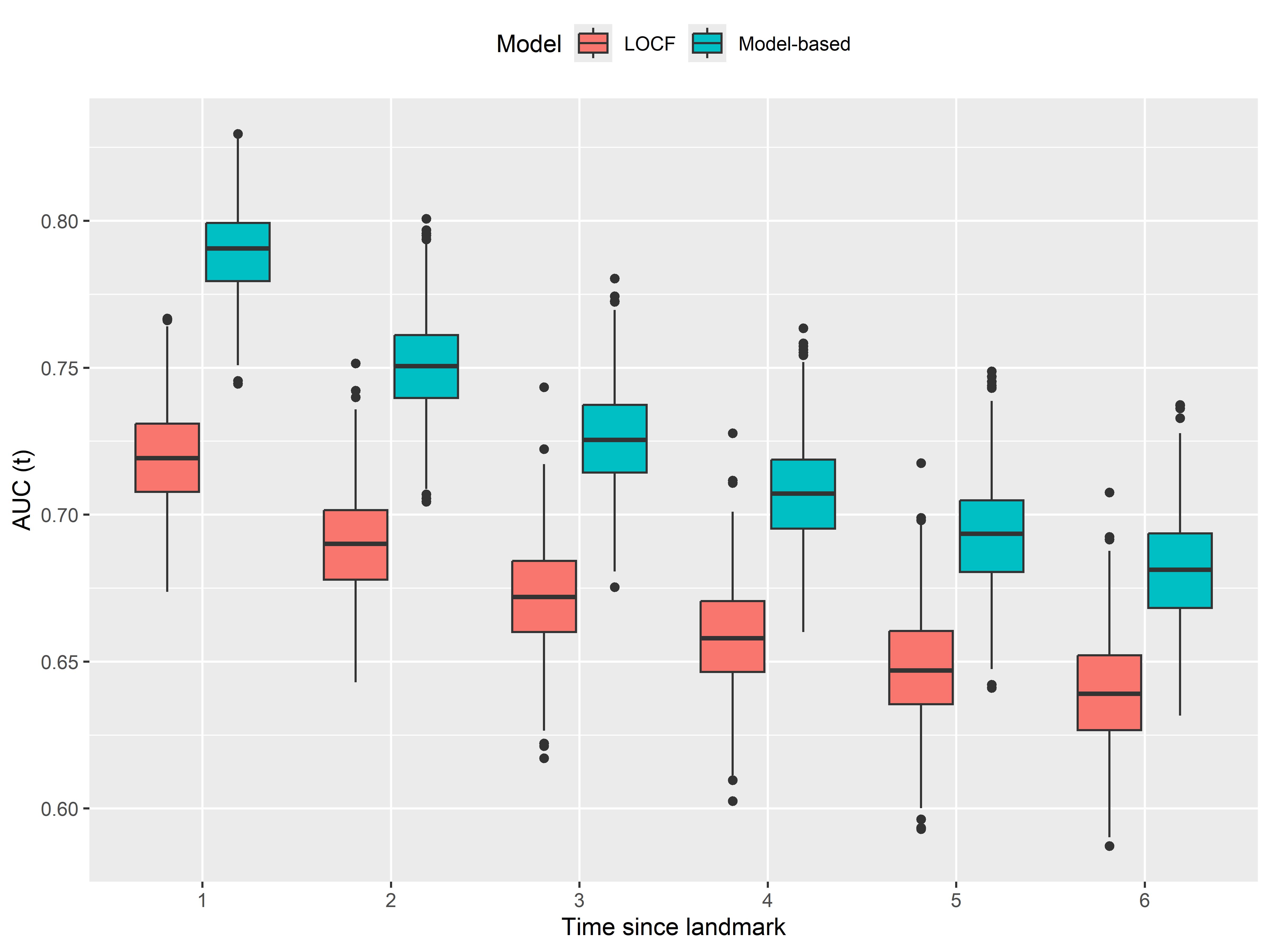}} 
\caption{Time-dependent AUC for scenarios under mild misspecification.}
\label{fig:mild}
\end{figure}

\begin{figure}[ht]
\centering
\textbf{$m=300$}\\
\subfloat[Scenario 9 - 20\% cured; $M$ balanced]{\includegraphics[width=.4\linewidth]{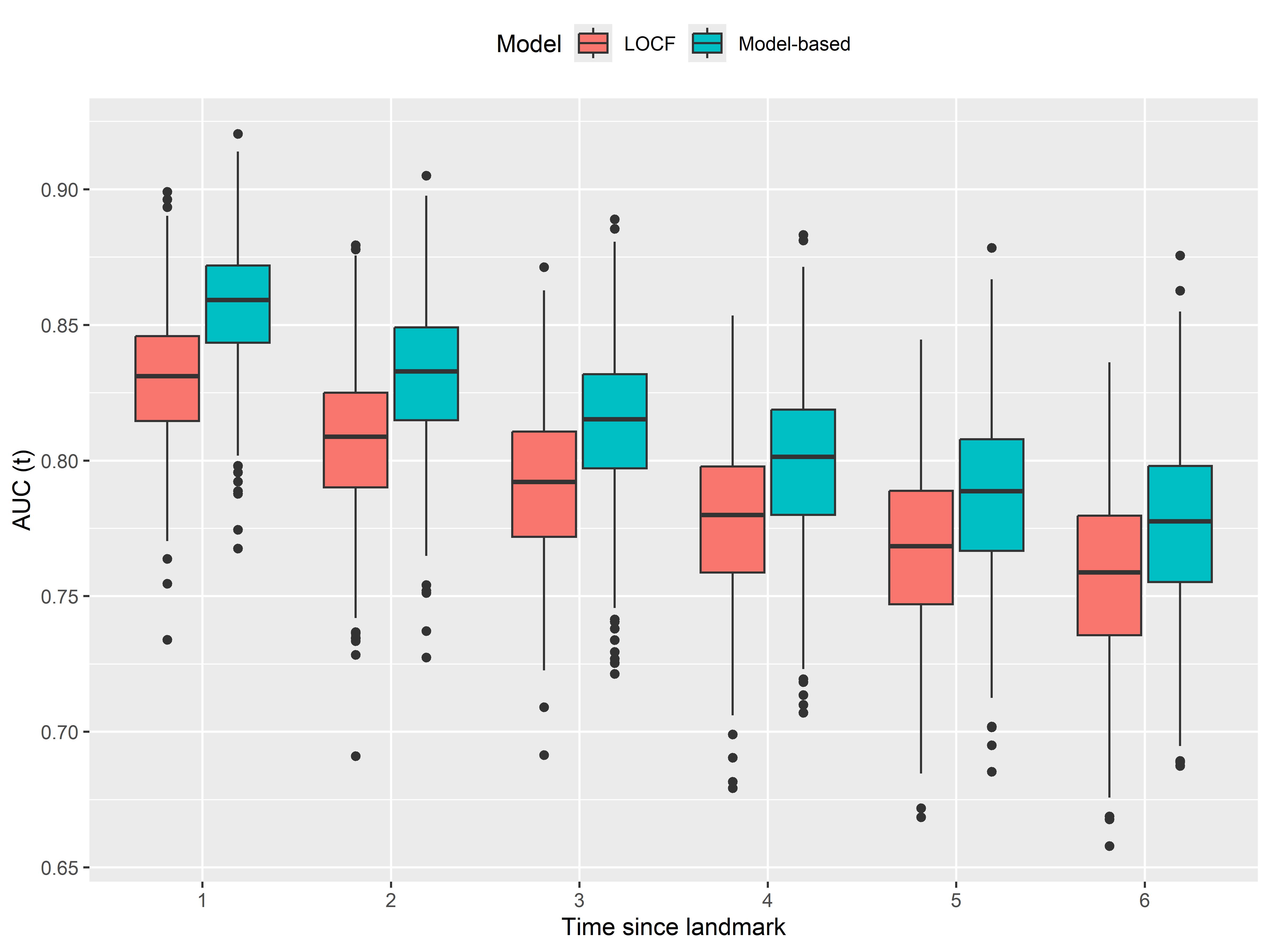}} 
\subfloat[Scenario 10 - 20\% cured; $M$ unbalanced]{\includegraphics[width=.4\linewidth]{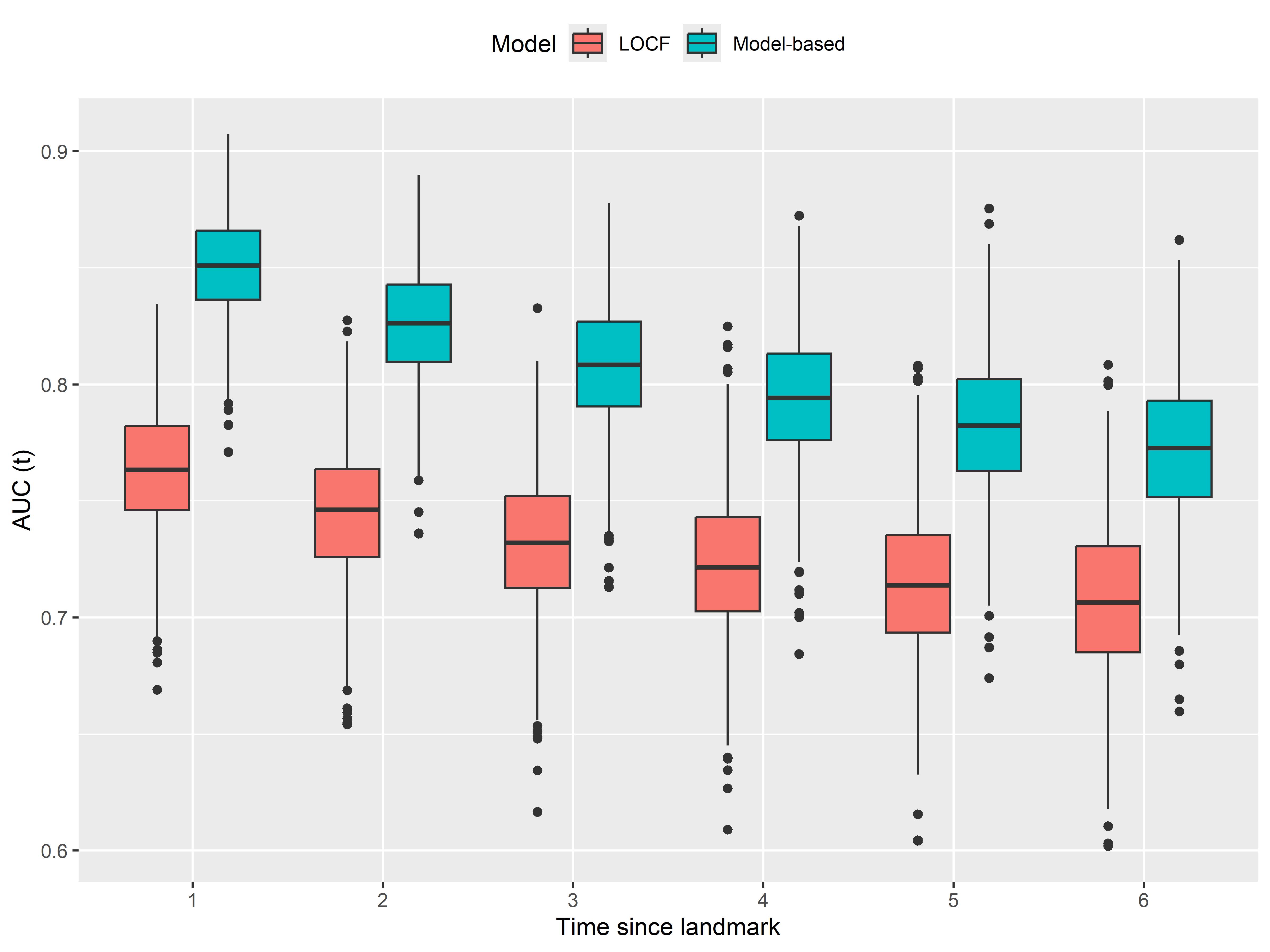}} \\
\subfloat[Scenario 11 - 40\% cured; $M$ balanced]{\includegraphics[width=.4\linewidth]{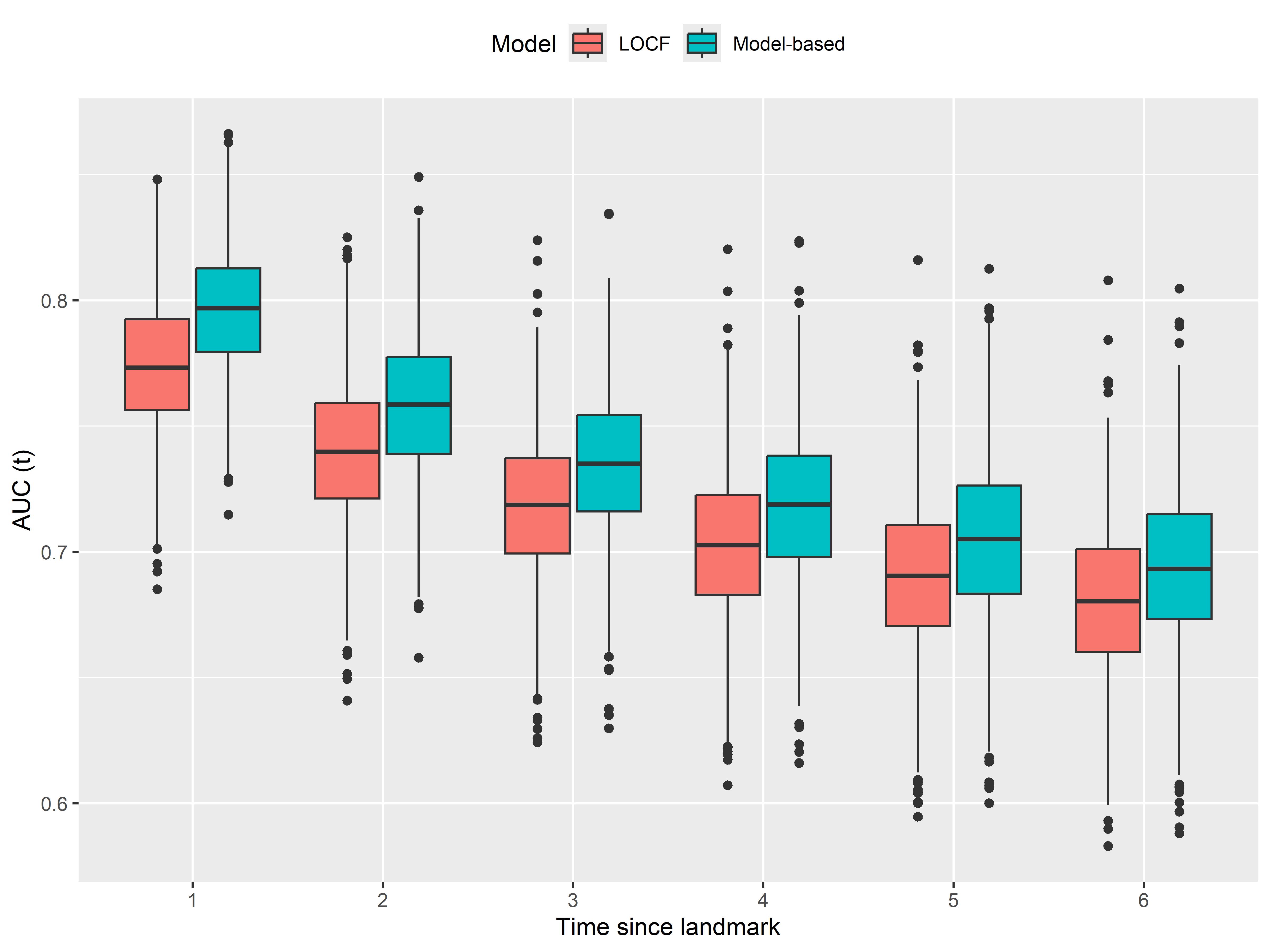}} 
\subfloat[Scenario 12 - 40\% cured; $M$ unbalanced]{\includegraphics[width=.4\linewidth]{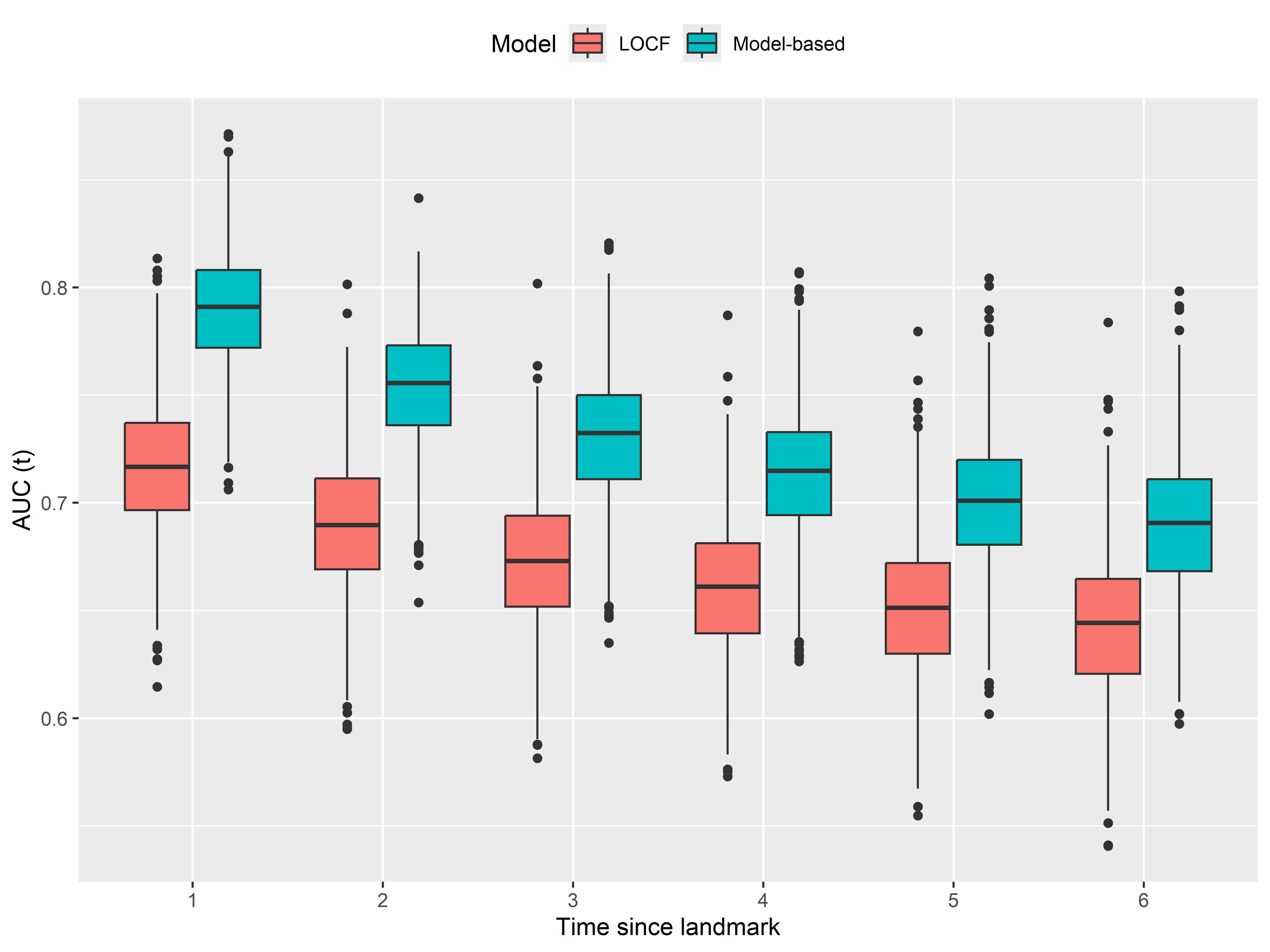}} \\[2mm]
\textbf{$m=1000$}\\
\subfloat[Scenario 9 - 20\% cured; $M$ balanced]{\includegraphics[width=.4\linewidth]{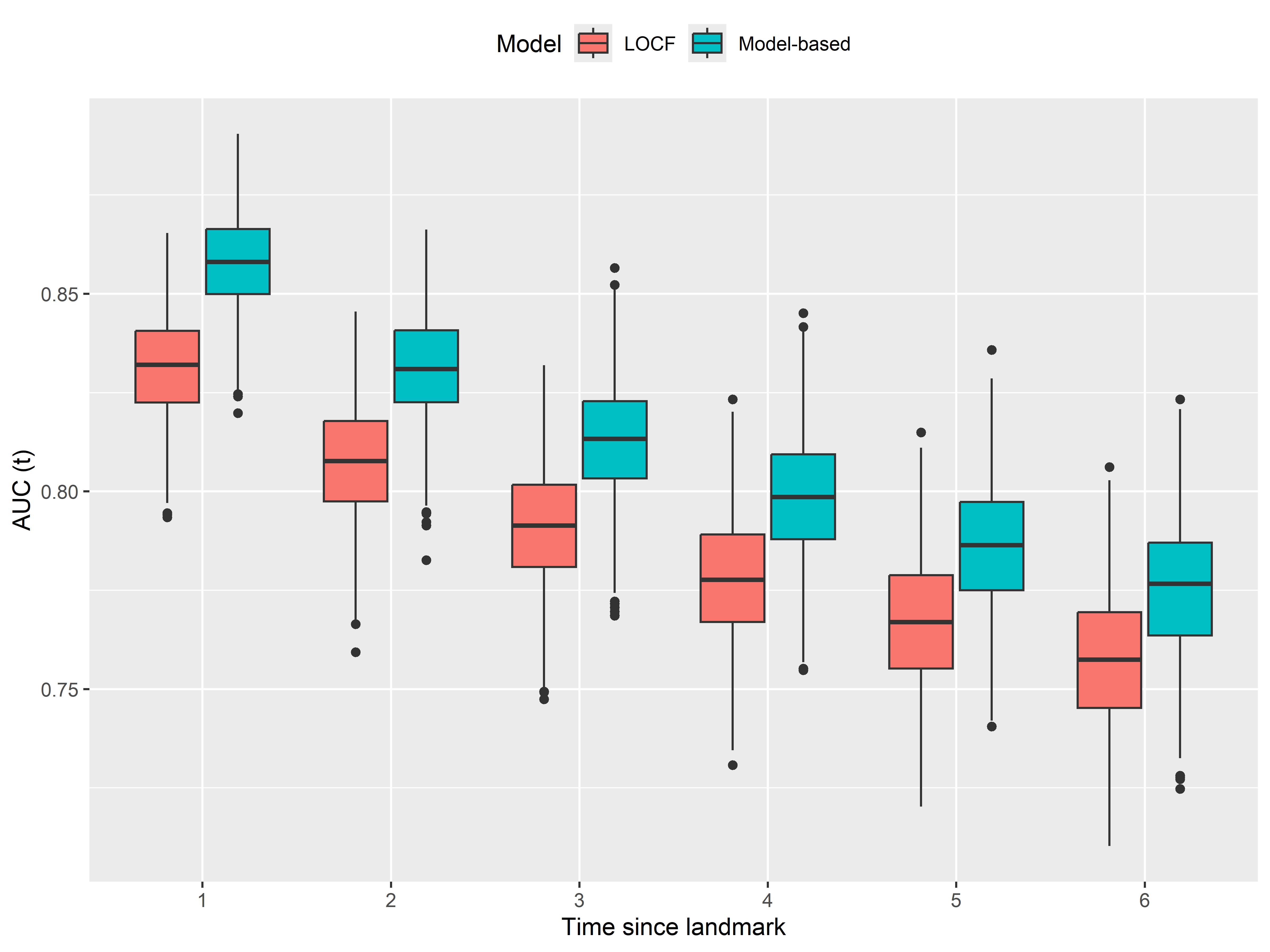}} 
\subfloat[Scenario 10 - 20\% cured; $M$ unbalanced]{\includegraphics[width=.4\linewidth]{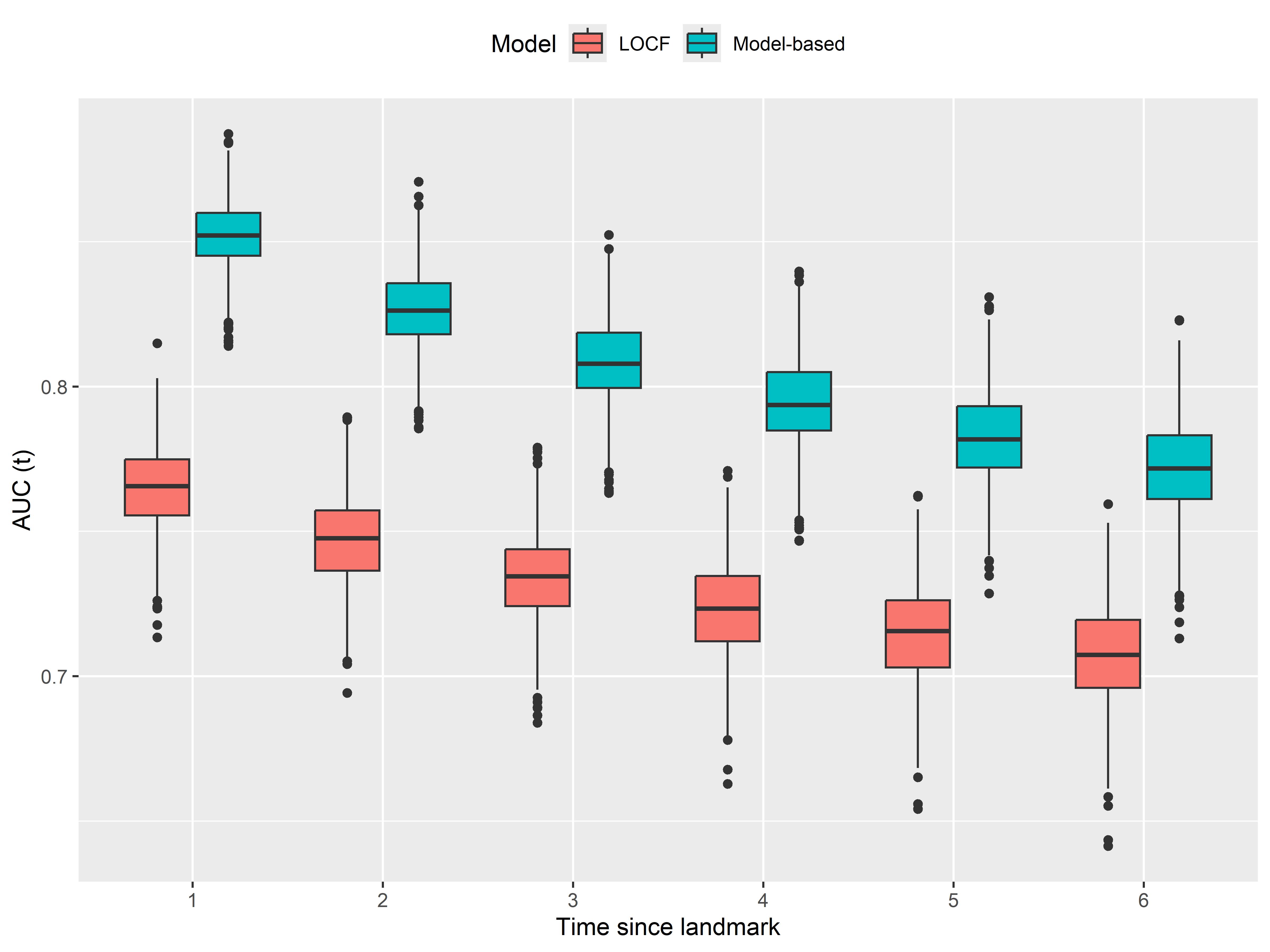}} \\
\subfloat[Scenario 11 - 40\% cured; $M$ balanced]{\includegraphics[width=.4\linewidth]{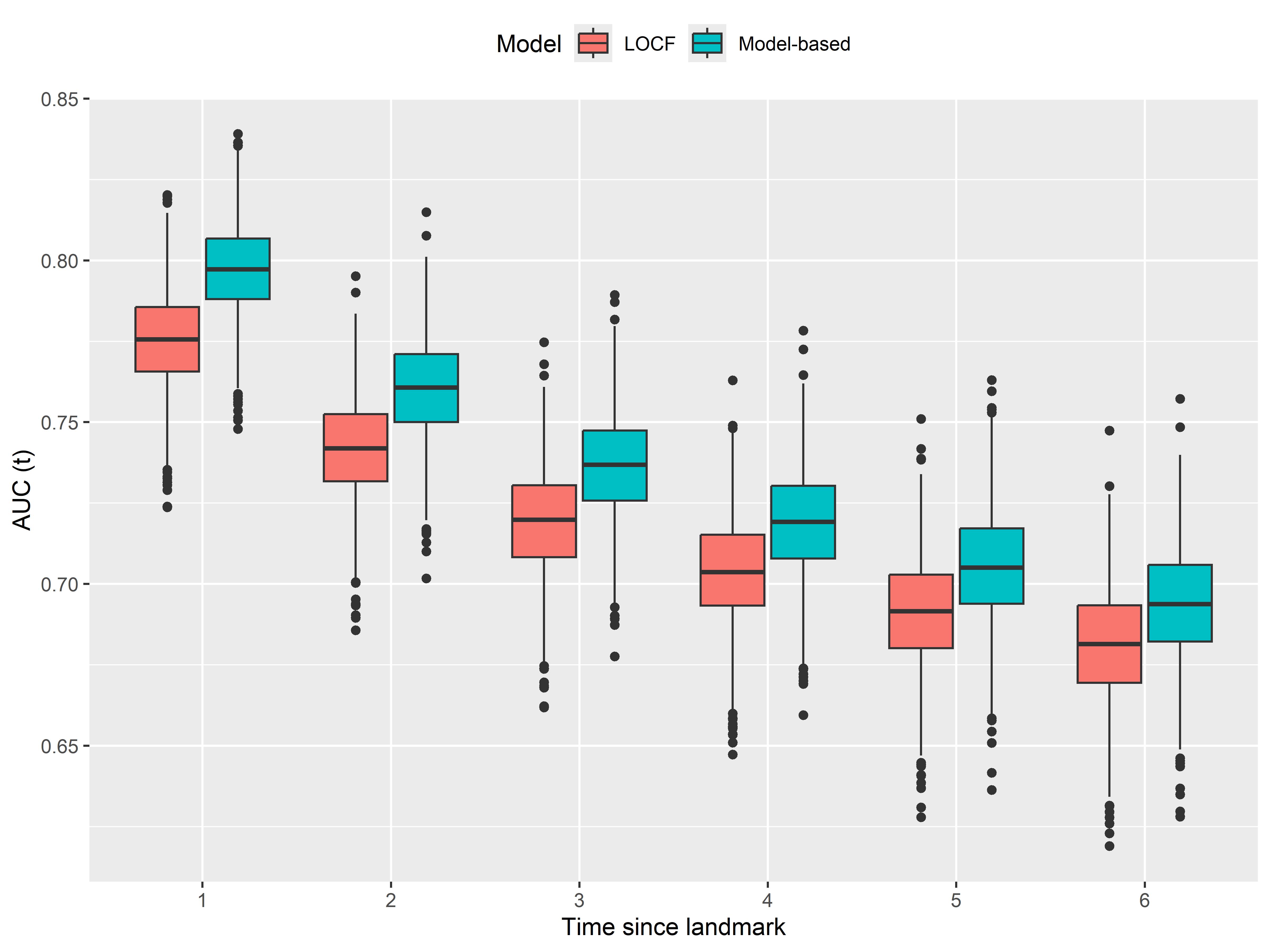}} 
\subfloat[Scenario 12 - 40\% cured; $M$ unbalanced]{\includegraphics[width=.4\linewidth]{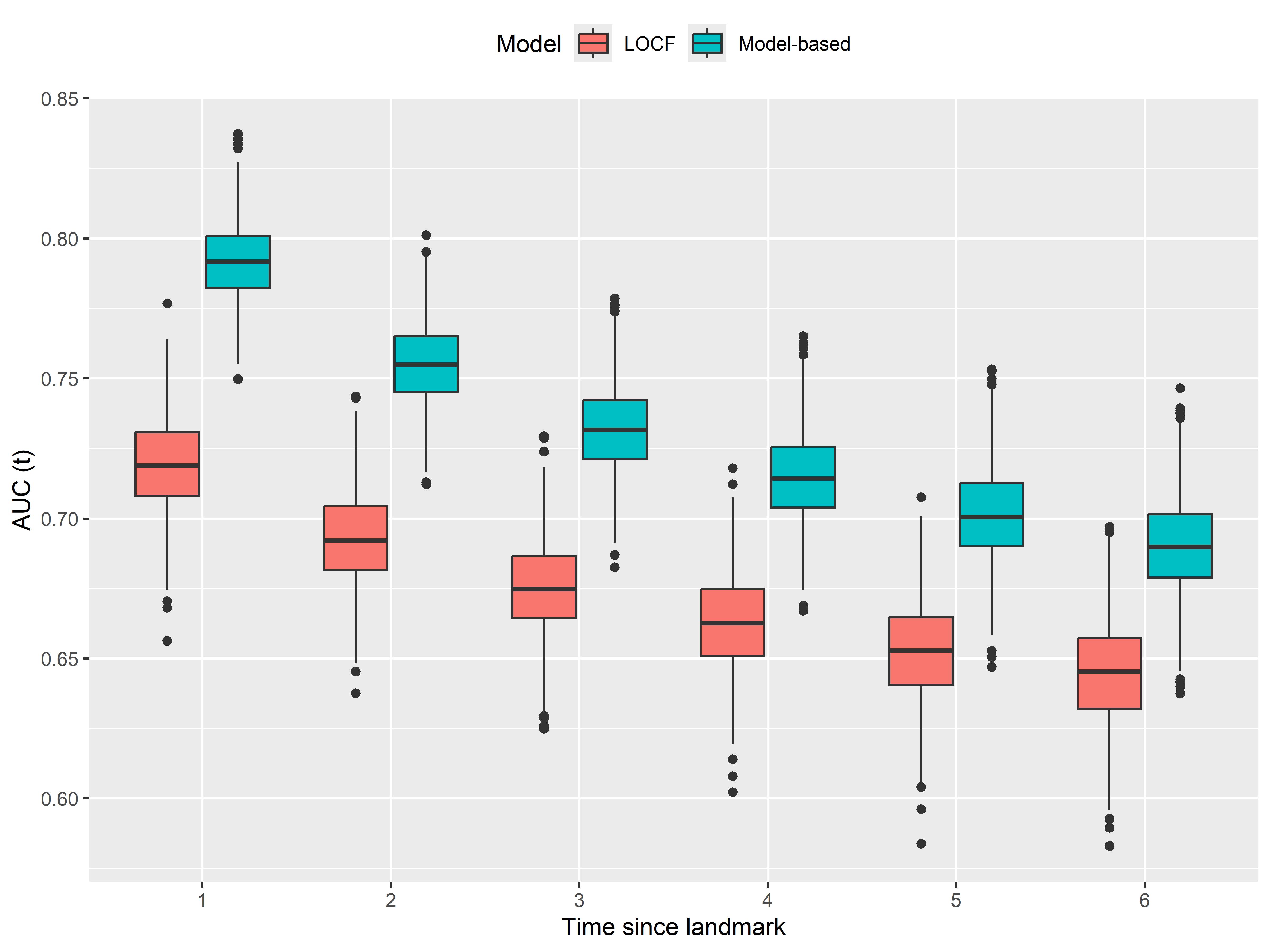}} 
\caption{Time-dependent AUC for scenarios under true data-generating process.}
\label{fig:true}
\end{figure}

Figures~\ref{fig:strong}, \ref{fig:mild}, and \ref{fig:true} show the time-dependent AUC for latency prediction under a strong misspecified, a mild misspecified, and a correctly specified model, respectively. 

Under strong misspecification (Figure~\ref{fig:strong}) and a balanced longitudinal study, the LOCF approach outperformed the model-based strategy. This finding is expected because, in this specific design, LOCF coincides with the true data-generating mechanism. However, when the longitudinal design was unbalanced, the proposed model-based approach outperformed the LOCF approach even though survival times were generated from the LOCF mechanism. This is explained by the fact that, due to irregular deletion, the LOCF estimation relied on measurements that could be substantially earlier than the landmark, effectively making the fitted LOCF model itself misspecified. This led to a marked loss of predictive performance, in line with the well-documented limitations of LOCF-based approaches under irregular observation patterns\cite{locf1, locf2}.

When moving to a mildly misspecified setting (Figure ~\ref{fig:mild}) or to a correctly specified setting (Figure ~\ref{fig:true}), the proposed model-based method consistently outperformed LOCF across all scenarios. This reflects the ability of the random-effects formulation to correctly capture individual trajectories when the data-generating process is at least approximately aligned with the assumed structure. Importantly, these conclusions were valid for both $n=300$ and $n=1000$, indicating that the comparative performance of the methods was not driven by the sample size. Additional performance metrics, reported in Figures 1-3 of the Supporting Information material, confirmed the patterns observed looking at the time-dependent AUC.

\section{Case study: renal transplant data}\label{sec:case}

\subsection{Data description and statistical modelling}
We illustrate the proposed methodology using the \texttt{renal} dataset from the \texttt{joineRML} R package\cite{joineRML}. This dataset comprises 407 patients with chronic kidney disease who received a primary renal transplant from either a deceased or living donor at the University Hospital of the Catholic University of Leuven (Belgium) between January 1983 and August 2000. Chronic kidney disease is characterized by progressive loss of renal function, typically assessed through repeated measurements of glomerular filtration rate (GFR).

\begin{figure}
    \centering
    \includegraphics[width=0.8\linewidth]{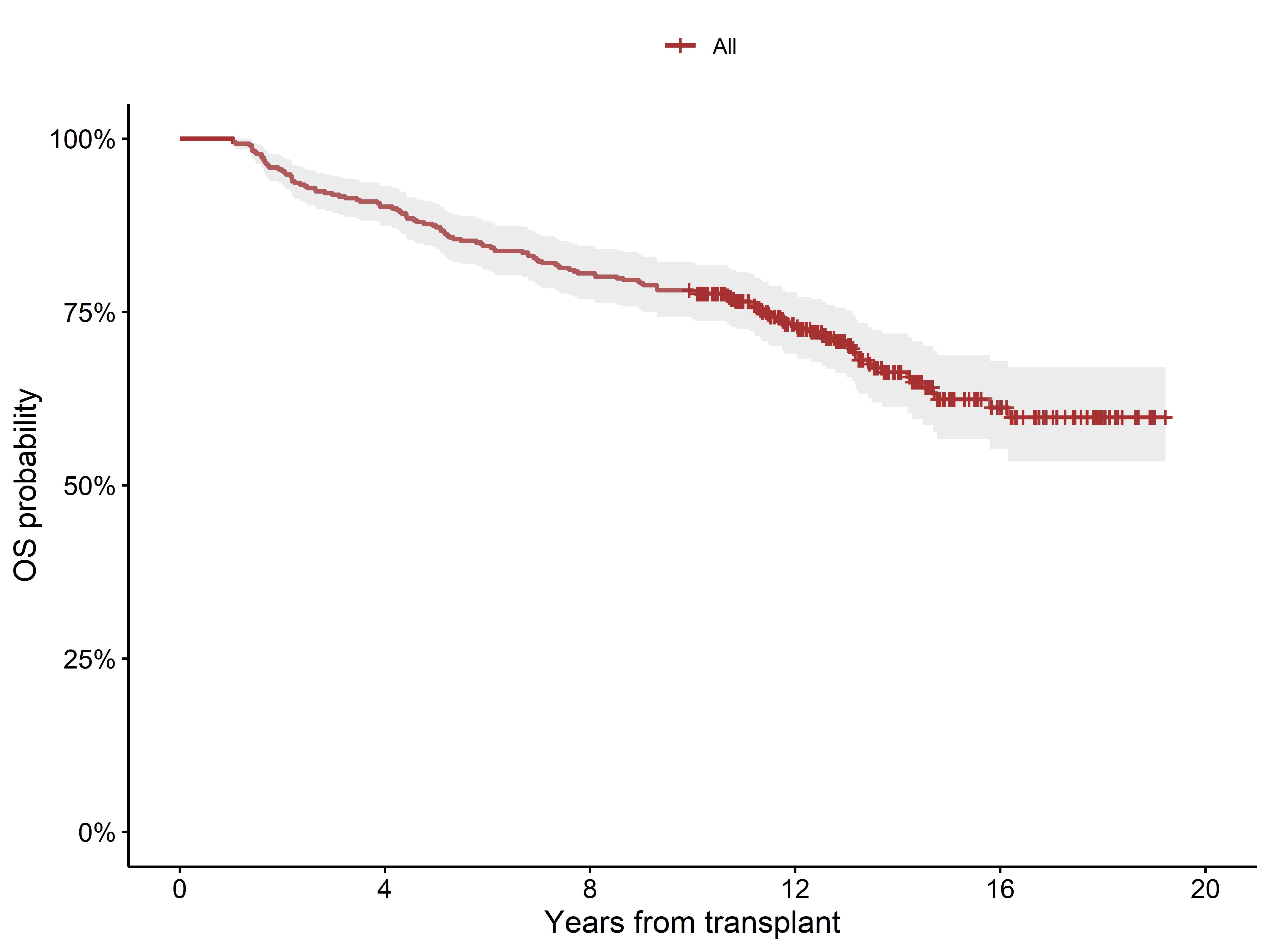}
    \caption{Kaplan-Meier survival curve for renal transplant patients}
    \label{fig:plateau}
\end{figure}

From a clinical perspective, renal transplantation is a potentially curative intervention, as successful transplantation restores kidney functions and long-term survival can be similar to that of the general population\cite{wolfe, lamb}. Accordingly, the Kaplan–Meier survival curve for this dataset (Figure~\ref{fig:plateau}) exhibits a plateau, a hallmark of cure scenarios\cite{amico}. 
In addition, such datasets have been widely used in methodological studies on joint modelling and dynamic prediction for renal transplant outcomes\cite{renal2, fieuws}, providing a rich framework for demonstrating the performance of advanced prediction techniques for graft failure.

For each patient, multiple longitudinal biomarkers were recorded over time, along with a time-to-event outcome indicating graft failure. We used baseline covariates, age, weight, and sex in the incidence submodel, with age categorized according to quartiles. Besides the marginal effect of age we also considered its interaction with sex. In the latency submodel, only longitudinal biomarkers reflecting kidney function were included, namely haematocrit (\% of red blood cells in blood) and GFR (ml/min/1.73 m$^2$).
For these two biomarkers, we report in Table \ref{tab:measures} the average number of measurements per individual at landmark times 1, 1.5, 2, 2.5 and 3 years after transplantation.

In the first step of the proposed method, each longitudinal biomarker was modeled using a linear mixed-effects model. To maintain computational simplicity and interpretability at the individual level, each model included only a random intercept and a random slope for time, allowing patient-specific deviations in both baseline levels and trajectories over time. This specification mirrors that used in the simulation study, without incorporating any additional model validation strategy.

\subsection{Results}
Like in our simulation study, we evaluated the performance of the model-based approach and compared it to the LOCF-based strategy. We used repeated cross validation with 4 folds and 10 repetitions to evaluate predicted performance.

It is important to note that the renal transplant dataset represents an atypical scenario: each patient has a very large number of repeated measurements (Table~\ref{tab:measures}), which can make modeling individual trajectories more challenging, while the abundance of data naturally favors LOCF. Despite these data characteristics, two distinct patterns emerged.  

\begin{figure}[ht]
\centering
\subfloat[Landmark time $t=1$]{\includegraphics[width=.5\linewidth]{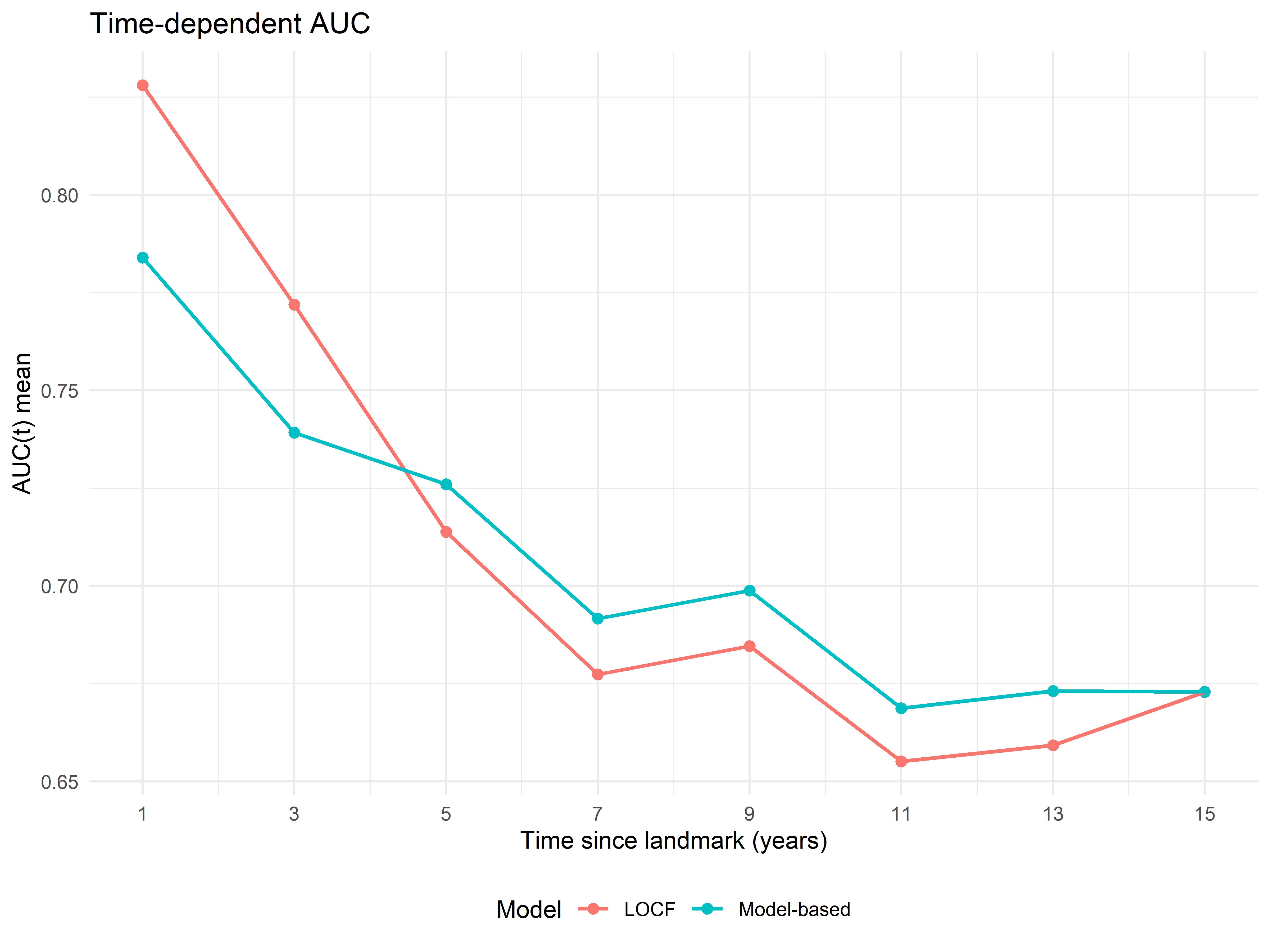}} 
\subfloat[Landmark time $t=1.5$]{\includegraphics[width=.5\linewidth]{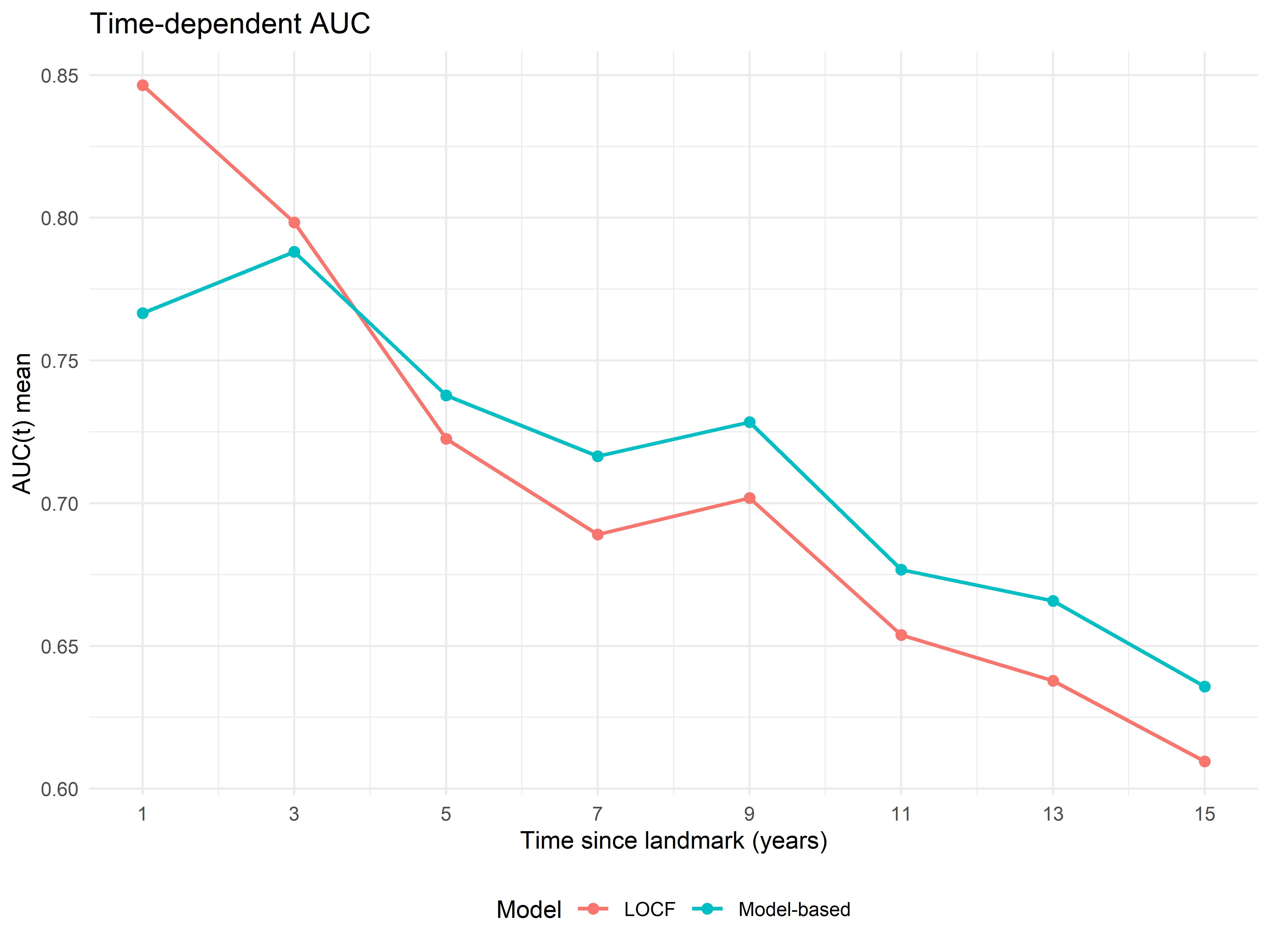}} \\
\subfloat[Landmark time $t=2$]{\includegraphics[width=.5\linewidth]{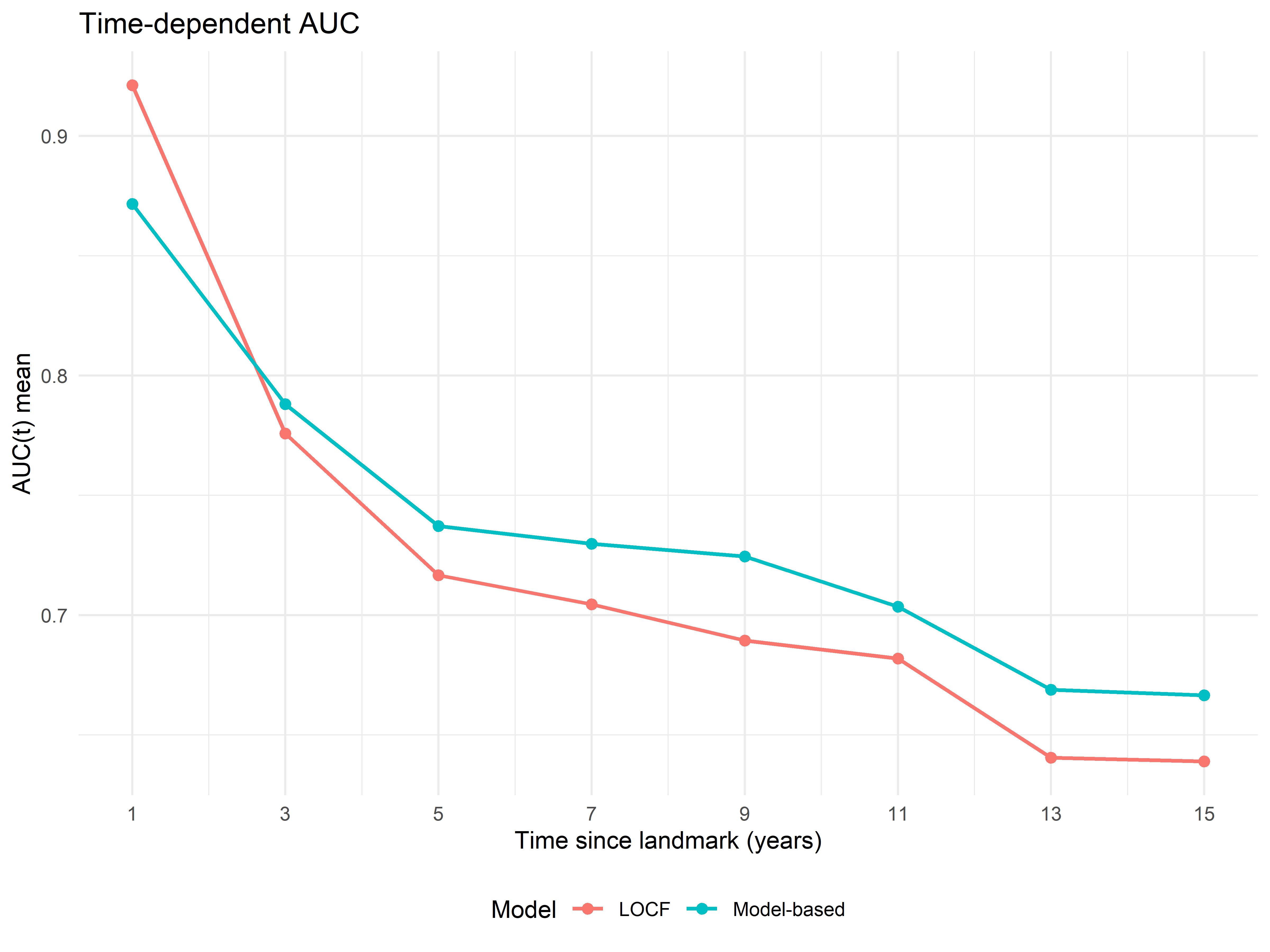}} 
\subfloat[Landmark time $t=2.5$]{\includegraphics[width=.5\linewidth]{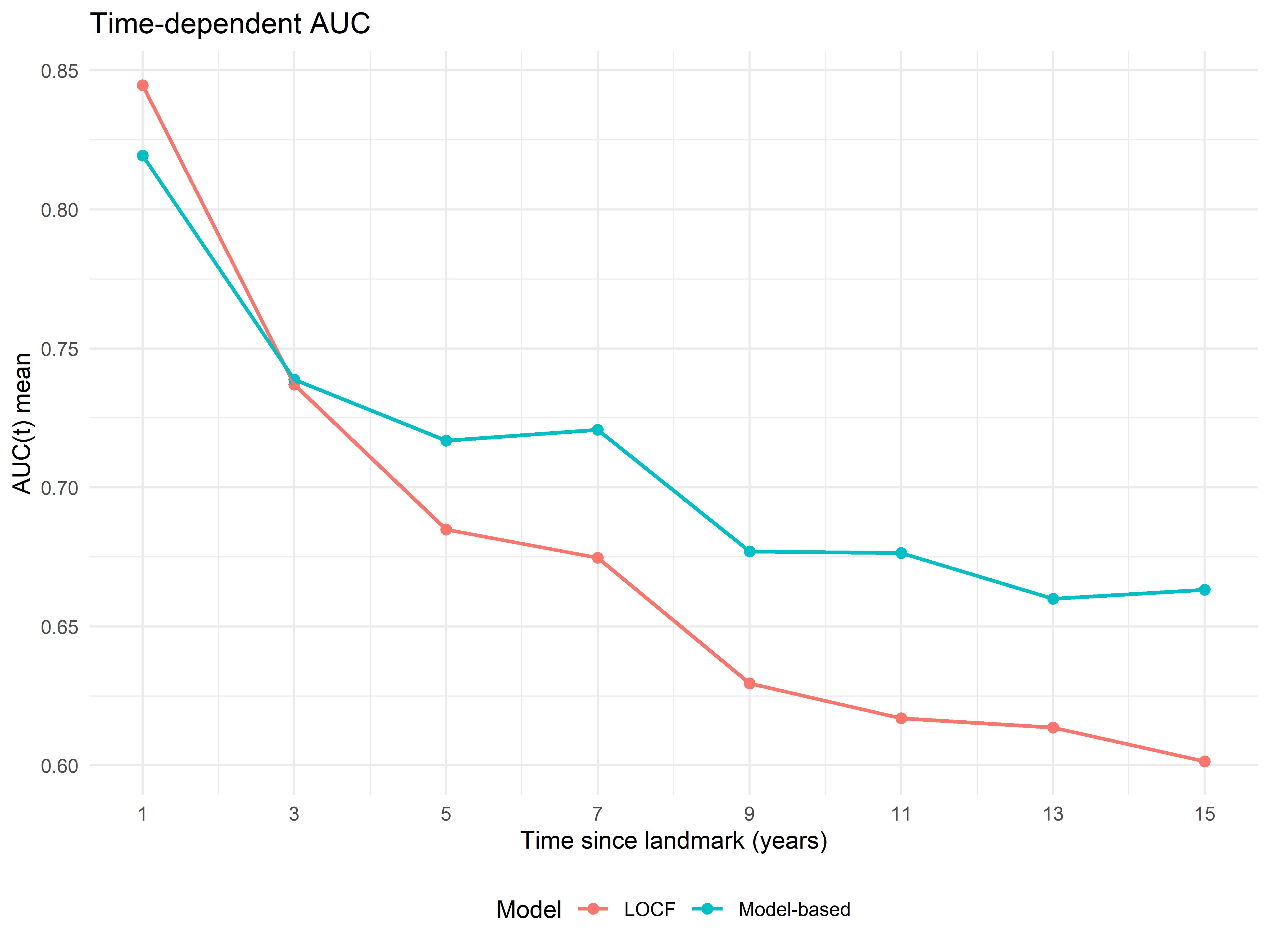}} \\
\subfloat[Landmark time $t=3$]{\includegraphics[width=.5\linewidth]{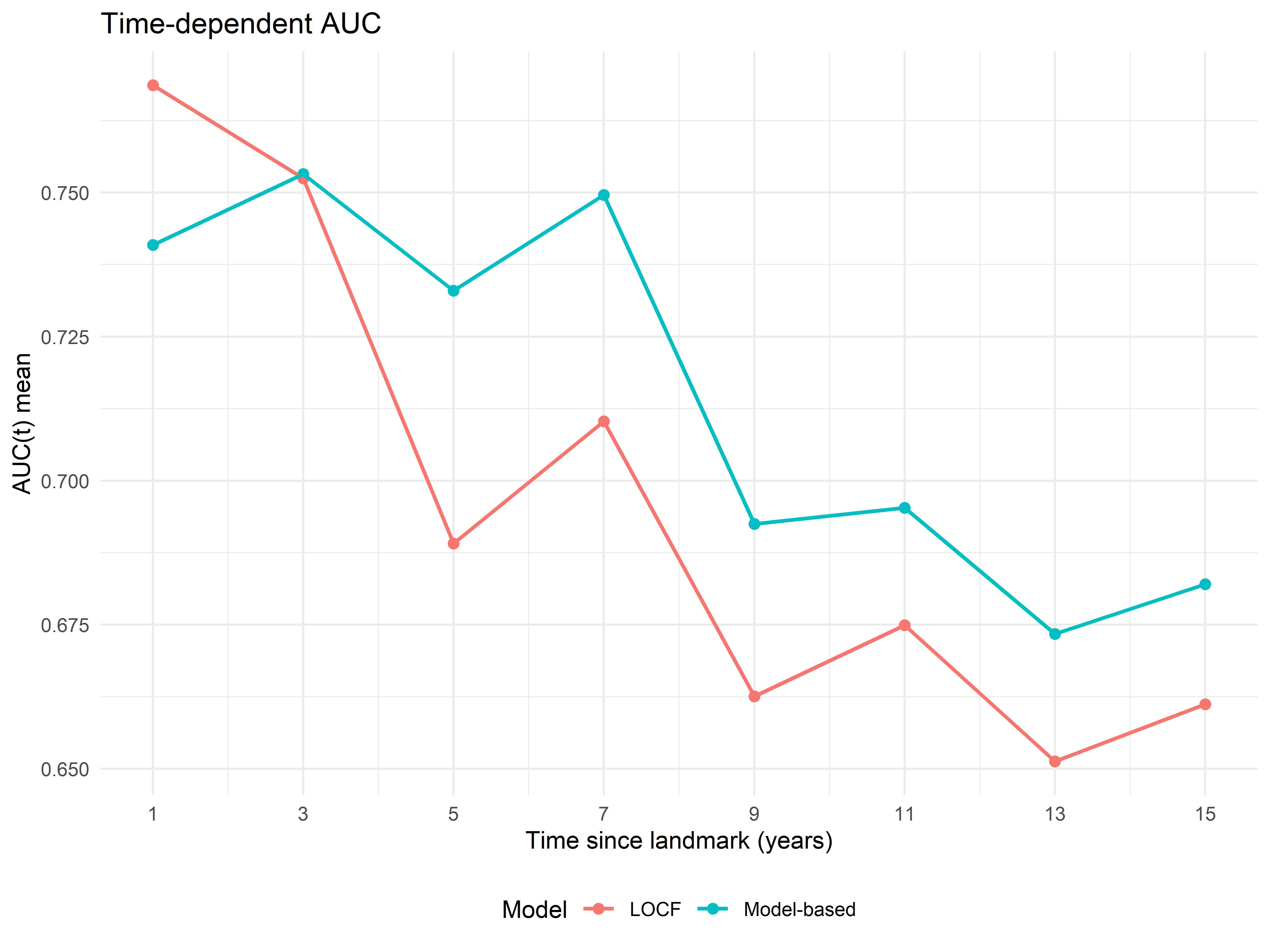}}  
\caption{Time-dependent AUC curves for the model-based and LOCF approaches at multiple landmark times.}
\label{fig:auc}
\end{figure}

First, the model-based approach shows an increasing relative advantage as the prediction horizon extends further from the fixed landmark time (typically after the first two years), as illustrated by the time-dependent AUC curves in Figure~\ref{fig:auc}. This highlights the benefit of explicitly modeling subject-specific trajectories rather than relying on the last available observation, which quickly loses relevance as time progresses.  

Second, when landmarks are set further from the baseline, the model-based approach also tends to perform better, since additional longitudinal information is available to inform the mixed-effects model. We can also notice that later landmarks are associated with fewer events available to estimate survival, leading to greater variability in performance metrics. 

Only AUC results are reported here; results based on the time-dependent Brier score, provided in Figure 5 of the Supporting Information material, showed trends consistent with the AUC but with smaller discrepancies between the two methods and reduced variability. In particular, both the evolution across landmark times and the performance decay when moving further from the landmark year appeared more stable with the Brier score. For clarity, Figure \ref{fig:auc} displays AUC curves without standard errors, while plots including standard errors for both the AUC and the Brier score are available in the Supporting Information material (Figures 4 and 6).

Table~\ref{tab:latency_cindex} reports the C-index for the latency submodel, providing a global measure of discriminative ability across landmarks. These values confirm the pattern observed with the AUC curves, with the model-based approach generally outperforming the LOCF-based one.  

For the incidence submodel, performance (Table~\ref{tab:incidence_metrics}) is broadly comparable between the model-based and LOCF-based approaches, consistent with the simulation study, reflecting that baseline covariates largely drive event occurrence predictions.


\section{Discussion}\label{sec:discussion}

In this work, we developed a methodology for dynamic prediction in the framework of mixture cure models with longitudinal covariates. Our approach is based on a model-based landmarking strategy in which we employ a mixed-effects model to obtain subject-specific summaries of the trajectories described by the longitudinal biomarkers. The method makes it possible to include longitudinal covariates into the latency component while maintaining computational tractability, making a step further in the literature on cure models, which so far has mainly focused on static covariates and LOCF-based landmarking.

The simulation study provided systematic evidence of the advantages of the model-based approach. Even when the data-generating mechanism coincided with LOCF, the proposed method performed competitively. Furthermore, once the longitudinal design became unbalanced, the proposed strategy showed clear superiority, reflecting the well-known limitations of LOCF-based approaches when measurement times are irregular. Under more realistic data-generating processes, whether mildly misspecified or correctly specified, the model-based approach consistently outperformed the LOCF-based one across all sample sizes, cure fractions, and longitudinal designs. These findings highlight the robustness of the proposed method in modelling misspecifications and its ability to leverage longitudinal information more effectively. Importantly, improvements were observed for both moderate and large sample sizes, suggesting that the benefits are not linked to the sample size.

The renal transplant case study further illustrated the practical relevance of our methodology. This dataset presents an atypical setting: patients undergo frequent follow-up measurements, leading to an abundance of longitudinal data that naturally favors the LOCF-based approach. Nevertheless, our results demonstrated that the model-based approach provides increasing predictive gains as the prediction horizon extends further from the landmark. This reflects its ability to model underlying trajectories rather than relying on a single past measurement, which becomes increasingly uninformative over time. At the same time, we observed the inherent trade-off of landmarking methods: while later landmarks benefit from richer longitudinal histories, they are associated with fewer events, which translates into greater variability of predictive metrics.

In both simulation and real data, a consistent pattern emerged between the incidence and latency components of the cure model. As the incidence component is primarily based on baseline covariates, the prediction performance for this component remained comparable between the two methods. In contrast, for the latency submodel, which directly incorporates longitudinal information, we observed substantial differences, with the model-based strategy showing clear advantages. This confirms the importance of explicitly modeling individual trajectories when the focus is on residual survival among the susceptible population.

From a methodological perspective, our study emphasizes the value of embedding longitudinal processes into cure models for prediction. While the LOCF-based approach remains attractive for its simplicity, it is prone to inaccuracy in settings with irregular follow-up and offers limited flexibility for extrapolating subject-specific trends. The proposed approach provides a principled alternative that strikes a balance between interpretability, computational feasibility, and predictive accuracy.

Despite its advantages, the proposed methodology also has some limitations. A key aspect of the model-based strategy is that it requires the specification of an appropriate mixed-effects model for the longitudinal trajectories in the first step. In practice, selecting the best-fitting structure (e.g., choice of random effects, covariance structure, or functional form of time) may not be straightforward, especially when data exhibit complex or highly irregular patterns. Misspecification at this stage can propagate to the prediction step, potentially reducing the accuracy of the latency component. Moreover, in some applications, simple linear time trends may be insufficient to capture the underlying biomarker dynamics, and more flexible models such as spline-based formulations or non-linear time effects may be necessary. While these extensions increase modeling accuracy, they also introduce additional challenges in terms of interpretation, computational burden, and risk of overfitting.

Future research may extend this framework in several directions. Incorporating multivariate longitudinal processes, for instance, through latent process mixed models, would allow joint modeling of multiple biomarkers that reflect a common biological pathway. Another avenue is to explore alternative parametrizations of the model, in which the summary of the longitudinal process used in the landmarked cure model is not the predicted random effects, but rather the predicted value of the longitudinal variable at the landmark time obtained from the mixed-effects model. 

Finally, we note that in the present work the landmark time was fixed and common to all individuals, defined in patient time. 
However, in other applications, the landmark time may arise naturally and be subject-specific, such as the time of transplantation for patients on a waiting list, after which survival prediction is of interest. 
In such settings, landmarking requires careful methodological adjustments because the length of the follow-up prior to the landmark differs across subjects. 
As shown by Gomon et al. (2023)\cite{gomon}, this may induce a selection bias in the estimation of both fixed and random effects of the longitudinal model, ultimately degrading predictive performance. 
Developing strategies, for instance, through appropriate reweighting, to account for individual-specific landmarks while preserving consistency of dynamic prediction in cure models represents an important direction for future research.

\bibliographystyle{vancouver}
\bibliography{main}

\clearpage
\section*{Tables}

\begin{table}[!htbp]
\centering
\caption{Simulation scenarios with scenario numbers. Each scenario is defined by cure fraction, longitudinal design, and generating mechanism.}
\label{tab:sim_scenarios}
\begin{tabular}{cccc}
\toprule
Scenario & Cure fraction & Longitudinal design & Generating mechanism \\
\midrule
1  & 20\% & Balanced ($M_i=10$)        & Strong misspecification (LOCF) \\
2  & 20\% & Unbalanced ($M_i \in [5,10]$) & Strong misspecification (LOCF) \\
3  & 40\% & Balanced ($M_i=10$)        & Strong misspecification (LOCF) \\
4  & 40\% & Unbalanced ($M_i \in [5,10]$) & Strong misspecification (LOCF) \\
5  & 20\% & Balanced ($M_i=10$)        & Mild misspecification (current value) \\
6  & 20\% & Unbalanced ($M_i \in [5,10]$) & Mild misspecification (current value) \\
7  & 40\% & Balanced ($M_i=10$)        & Mild misspecification (current value) \\
8  & 40\% & Unbalanced ($M_i \in [5,10]$) & Mild misspecification (current value) \\
9  & 20\% & Balanced ($M_i=10$)        & True model (random effects) \\
10 & 20\% & Unbalanced ($M_i \in [5,10]$) & True model (random effects) \\
11 & 40\% & Balanced ($M_i=10$)        & True model (random effects) \\
12 & 40\% & Unbalanced ($M_i \in [5,10]$) & True model (random effects) \\
\bottomrule
\end{tabular}
\end{table}

\clearpage

\begin{table}[!htbp]
\centering
\caption{Incidence results for each scenario (mean [SD]) by sample size. Scenario numbers refer to Table~\ref{tab:sim_scenarios}. Results are reported for LOCF and model-based approaches, with weighted AUC and weighted Brier score.}
\label{tab:incidence_results_n300_n1000}
\begin{tabular}{ccc}
\toprule
\multicolumn{3}{c}{\textbf{Sample size $n=300$}} \\
\midrule
Scenario & LOCF (AUC [SD]; Brier [SD]) & Model-based (AUC [SD]; Brier [SD]) \\
\midrule
1  & 0.75 [0.04]; 0.14 [0.01] & 0.75 [0.04]; 0.15 [0.01] \\
2  & 0.75 [0.04]; 0.15 [0.01] & 0.75 [0.04]; 0.15 [0.01] \\
3  & 0.77 [0.03]; 0.19 [0.01] & 0.77 [0.03]; 0.19 [0.01] \\
4  & 0.77 [0.03]; 0.19 [0.01] & 0.77 [0.03]; 0.19 [0.01] \\
5  & 0.75 [0.04]; 0.14 [0.01] & 0.75 [0.04]; 0.14 [0.01] \\
6  & 0.75 [0.03]; 0.14 [0.01] & 0.75 [0.03]; 0.14 [0.01] \\
7  & 0.77 [0.03]; 0.19 [0.01] & 0.77 [0.03]; 0.19 [0.01] \\
8  & 0.77 [0.03]; 0.19 [0.01] & 0.77 [0.03]; 0.19 [0.01] \\
9  & 0.75 [0.04]; 0.15 [0.01] & 0.75 [0.04]; 0.15 [0.01] \\
10 & 0.75 [0.04]; 0.15 [0.01] & 0.75 [0.04]; 0.14 [0.01] \\
11 & 0.77 [0.03]; 0.19 [0.01] & 0.77 [0.03]; 0.19 [0.01] \\
12 & 0.77 [0.03]; 0.19 [0.01] & 0.77 [0.03]; 0.19 [0.01] \\
\midrule
\multicolumn{3}{c}{\textbf{Sample size $n=1000$}} \\
\midrule
Scenario & LOCF (AUC [SD]; Brier [SD]) & Model-based (AUC [SD]; Brier [SD]) \\
\midrule
1  & 0.76 [0.02]; 0.14 [0.01] & 0.76 [0.02]; 0.14 [0.01] \\
2  & 0.76 [0.02]; 0.15 [0.01] & 0.76 [0.02]; 0.14 [0.01] \\
3  & 0.78 [0.01]; 0.19 [0.01] & 0.78 [0.01]; 0.19 [0.01] \\
4  & 0.78 [0.01]; 0.19 [0.01] & 0.78 [0.01]; 0.19 [0.01] \\
5  & 0.76 [0.02]; 0.14 [0.01] & 0.76 [0.02]; 0.14 [0.01] \\
6  & 0.76 [0.02]; 0.14 [0.01] & 0.76 [0.02]; 0.14 [0.01] \\
7  & 0.78 [0.01]; 0.19 [0.01] & 0.78 [0.01]; 0.19 [0.01] \\
8  & 0.78 [0.01]; 0.19 [0.01] & 0.78 [0.01]; 0.19 [0.01] \\
9  & 0.76 [0.02]; 0.14 [0.01] & 0.76 [0.02]; 0.14 [0.01] \\
10 & 0.76 [0.02]; 0.15 [0.01] & 0.76 [0.02]; 0.14 [0.01] \\
11 & 0.77 [0.01]; 0.19 [0.01] & 0.77 [0.01]; 0.19 [0.01] \\
12 & 0.78 [0.01]; 0.19 [0.01] & 0.78 [0.01]; 0.19 [0.01] \\
\bottomrule
\end{tabular}
\end{table}

\clearpage

\begin{table}[!htbp]
\centering
\caption{Average number of longitudinal measurements per patient at selected landmark times}
\label{tab:measures}
\begin{tabular}{c|cc}
\hline
\textbf{Landmark time (years)} & \textbf{Haematocrit} & \textbf{GFR} \\
\hline
1.0   & 54.07 & 59.25 \\
1.5   & 61.55 & 68.44 \\
2.0   & 67.31 & 75.96 \\
2.5   & 72.69 & 82.39 \\
3.0   & 77.56 & 87.93 \\
\hline
\end{tabular}
\end{table}

\clearpage

\begin{table}[!htbp]
\centering
\caption{Incidence metrics (AUC and Brier) at different landmark times for Model-based and LOCF approaches.}
\label{tab:incidence_metrics}
\begin{tabular}{cccccc}
\hline
Landmark time (years) & Model & AUC Mean & AUC SD & Brier Mean & Brier SD \\
\hline
1 & Model-based & 0.76 & 0.03 & 0.23 & 0.03 \\
1 & LOCF & 0.76 & 0.03 & 0.23 & 0.03 \\
1.5 & Model-based & 0.74 & 0.03 & 0.23 & 0.03 \\
1.5 & LOCF & 0.75 & 0.03 & 0.24 & 0.03 \\
2 & Model-based & 0.77 & 0.03 & 0.22 & 0.03 \\
2 & LOCF & 0.77 & 0.03 & 0.23 & 0.03 \\
2.5 & Model-based & 0.78 & 0.03 & 0.22 & 0.03 \\
2.5 & LOCF & 0.79 & 0.03 & 0.22 & 0.04 \\
3 & Model-based & 0.77 & 0.03 & 0.21 & 0.03 \\
3 & LOCF & 0.78 & 0.03 & 0.22 & 0.03 \\
\hline
\end{tabular}
\end{table}

\clearpage

\begin{table}[!htbp]
\centering
\caption{Latency metrics (C-index) at different landmark times (years) for Model-based and LOCF approaches.}
\label{tab:latency_cindex}
\begin{tabular}{cccc}
\hline
Landmark Time (years) & Model       & C-index Mean & C-index SD \\
\hline
1   & Model-based & 0.66 & 0.03 \\
1   & LOCF        & 0.66 & 0.05 \\
1.5 & Model-based & 0.68 & 0.05 \\
1.5 & LOCF        & 0.65 & 0.05 \\
2   & Model-based & 0.69 & 0.06 \\
2   & LOCF        & 0.67 & 0.06 \\
2.5 & Model-based & 0.68 & 0.06 \\
2.5 & LOCF        & 0.64 & 0.07 \\
3   & Model-based & 0.69 & 0.06 \\
3   & LOCF        & 0.67 & 0.06 \\
\hline
\end{tabular}
\end{table}

\end{document}